\documentclass[letterpaper,twocolumn,10pt,]{article}
\usepackage{usenix2019_v3}

\usepackage{tikz}
\usepackage{amsmath}
\usepackage{amsthm}

\usepackage{filecontents}

\usepackage[english]{babel}
\usepackage{blindtext}
\usepackage{xcolor}
\usepackage{gincltex}
\usepackage{pgfplots, siunitx, tikzscale}
\usepackage{graphicx}
\usepackage{subfigure}
\usepackage{pifont}
\usetikzlibrary{pgfplots.statistics}
\usepackage{float}
\usepackage{tcolorbox}
\usepackage { bb m}
\usepackage{enumitem}
\usepackage{booktabs}
\usepackage{makecell}
\usepackage {pifont}
\usepackage{amssymb}
\usepackage{algorithm}
\usepackage{algpseudocode}
\usepackage{inconsolata}
\usepackage{tabularx}
\usepackage{multirow}
\usepackage{multicol}
\usepackage[font=bf, size=small]{caption}
\newcommand{\name}{RegenHance\xspace}
\newcommand{\hide}[1] {}

\definecolor{mypink1}{rgb}{0.858, 0.188, 0.478}

\vspace{0.4em}
\newcommand{\ie}{\emph{i.e.}\xspace}
\newcommand{\eg}{\emph{e.g.}\xspace}
\newcommand{\vs}{\emph{v.s.}\xspace}

\usepackage{titlesec}
\titleformat{\subsection}
  {\normalfont\fontsize{10.5}{12}\bfseries}{\thesubsection}{1em}{}
\titleformat{\subsubsection}[runin]{\normalfont\bfseries}{\thesubsubsection}{1em}{}
\titlespacing*{\subsubsection}{0pt}{\dimexpr 1ex minus 0.7ex}{\dimexpr 1ex plus 0.7ex}

\title{\Large \bf Region-based Content Enhancement for Efficient Video Analytics at the Edge}

 \author{
{\rm Weijun Wang$^1$\footnotemark[1]\; Liang Mi$^2$\footnotemark[1]\,\,\footnotemark[2]\; Shaowei Cen$^2$\footnotemark[2]\; Haipeng Dai$^2$\; Yuanchun Li$^{1}$\; Xiaoming Fu$^3$\; Yunxin Liu$^{1}$\footnotemark[3]}\\
$^1$Institute for AI Industry Research (AIR), Tsinghua University\\
$^2$State Key Laboratory for Novel Software Technology, Nanjing University\\
$^3$Institute of Computer Science, University of G\"ottingen
 } 
\begin{document}

\maketitle

\renewcommand{\thefootnote}{\fnsymbol{footnote}} 
\footnotetext[1]{Weijun Wang and Liang Mi contributed equally to this work.} 
\footnotetext[2]{This work was done while Liang Mi and Shaowei Cen were interns at the Institute for AI Industry
Research (AIR), Tsinghua University.} 
\footnotetext[3]{Corresponding author: Yunxin Liu <liuyunxin@air.tsinghua.edu.cn>.}





\begin{abstract}
Video analytics is widespread in various applications serving our society.
Recent advances of content enhancement in video analytics offer significant benefits for bandwidth saving and accuracy improvement. 
However, existing content-enhanced video analytics systems are prohibitively computationally expensive and provide extremely low throughput.
In this paper, we present region-based content enhancement, 
that enhances only the important regions in videos, to improve analytical accuracy.
Our system, \name, enables high-accuracy and high-throughput video analytics at the edge by 1) a macroblock-based region importance predictor that identifies the important regions fast and precisely, 2) a region-aware enhancer 
that stitches sparsely distributed regions into dense tensors and enhances them efficiently,
and 3) a profile-based execution planer that allocates appropriate resources for enhancement and analytics components.
We prototype \name on five heterogeneous edge devices.
Experiments on two analytical tasks reveal that region-based enhancement improves the overall accuracy of 10-19\% and achieves 2-3$\times$ throughput compared to the state-of-the-art frame-based enhancement methods. 
\end{abstract}
\section{Introduction}
Video cameras are widespread in our society, with numerous installations in major cities and organizations \cite{Tiananmen, Chicago, Britain, British, VideoCovid}.
These cameras continuously collect vast amounts of video data 
for various applications such as traffic control~\cite{Ganesh2017Real}, school security~\cite{datondji2016survey}, crime investigation~\cite{wang2021Poster, Retail}, sports refereeing~\cite{Sports, Qatar}, and more \cite{Change, wolk2021measuring}.
Significant advances in deep neural networks (DNNs) for vision processing offer a tremendous opportunity to AI-powered automatic \emph{video analytics}~\cite{jiang2018chameleon, zhang2018awstream, xie2019source, du2022AccMPEG, du2020server, ye2023accelir}. 
A typical video analytics pipeline follows the distributed paradigm as AI methods often require high computing power that cameras cannot offer \cite{WyzeCamera, AXISCamera}.
Cameras capture and deliver the live videos to the edge in proximity for real-time processing.
However, the outdated hardware of cameras in use \cite{li2020reducto} and extremely limited uplink bandwidth between the camera and the edge \cite{speedtest, cell, zhang2018awstream} results in limited video quality and thus low analytical accuracy.




\emph{Content enhancement} offers a promising solution to tackle this issue.
It delivers a remarkable accuracy improvement \cite{yi2020eagleeye, wang2022enabling, lu2022turbo, AccDecoder, wang2020dual, noh2019better, tan2018feature} and bandwidth saving \cite{yi2020supremo, yeo2018neural, yeo2020nemo, kim2020neural, sun2024biswift, yeo2022neuroscaler, dasari2020streaming} by leveraging neural enhancement models (\eg, super-resolution \cite{ahn2018fast, Lim2017EDSR}, generative adversarial network \cite{bulat2018super, gulrajani2017improved}, image restoration \cite{liang2021swinir}) to enhance the informative details of video frames prior to feeding them into the final analytical models. 
Different from previous model optimization methods (\eg, model merging~\cite{padmanabhan2022gemel, Jiang2018Mainstream}, model updating~\cite{Romil2022Ekya, kim2020neural}, and model switching~\cite{khani2023recl, yeo2022neuroscaler}), data-optimization content enhancement does not require modifying user-provide models~\cite{agarwal2023boggart, poddar2020visor, cangialosi2022privid}.
Besides, even if the city government updates current low-end cameras to the latest ones that can offer high-quality video, content enhancement can still improve the details of small objects or blurred content.




%

Unfortunately, naively employing content enhancement in practice is excessively computationally expensive.
Such straightforward ways of enhancements not only cause high latency but also compete for computational resources with final analytical models. 
For example, applying enhancement on one tail-accuracy frame or executing generative adversarial networks on hard-recognized human faces causes hundreds to thousands of milliseconds of latency~\cite{wang2022enabling, yi2020eagleeye}.
The state-of-the-art method, \emph{selective enhancement}~\cite{yeo2020nemo, yeo2022neuroscaler}, yields some throughput improvement by enhancing only sampled frames but results in substantial accuracy reduction (more in \S~\ref{sec:bottleneck}).


\vspace{0.3em}\noindent
\textbf{Goal and observations.} 
This paper asks the following research question: Can we integrate content enhancement into video analytics at the edge in a high-throughput manner?
We argue that yes, but it requires a rethink of enhancement.
Our key observations are that 1) the time cost of enhancement positively correlates to the input size, and 2) the regions after enhancement that benefit analytical accuracy only occupy a small portion in each frame (more details in \S \ref{sec:OurChoice}).
Therefore, we aim to develop a \emph{region-based content-enhancement} method which only enhances the beneficial regions.

To this goal, we face three non-trivial challenges.

\textbf{(C1) How to identify beneficial regions fast and accurately?} 
Simply relying on DNN-based identification is too slow to the extent that it even exceeds the time required for the entire frame enhancement.

\textbf{(C2) How to efficiently enhance regions?} 
Characteristics of enhancement and heterogeneous accuracy gain of regions necessitate a novel enhancement mechanism that differs from the previous frame-based enhancement \cite{yeo2020nemo, yeo2018neural, kim2020neural, yeo2022neuroscaler}.

\textbf{(C3) How to allocate resources among components?} 
To maximize the overall performance, 
limited resources on edge must be 
best allocated among system components; 
however, typical schedulers cause 
imbalances 
among components, which in turn leads to noticeable throughput degradation.


\noindent
\textbf{\name.}
This paper presents \name, a system that efficiently identifies and enhances the most beneficial regions in video, enabling high-accuracy and high-throughput analytics at the edge.
\name addresses the above challenges by following techniques:



\emph{Macroblock-based region importance prediction.} 
We propose a predictor to fast and accurately identify the regions that yield the highest accuracy improvements in videos (\S \ref{sec:RegionSelector}).
We first analyze the advantages of predicting the region importance at the macroblock (MB, the video encoding unit) level and establish a metric to precisely measure the accuracy gain (importance) of each MB after enhancement in each frame.  
Next, we 
develop an ultra-lightweight prediction model and a prediction results reuse algorithm, achieving high throughput. 
Experiments show our method can predict the MB importance in 30 frames per second on a single CPU thread.

\emph{Region-aware enhancement.} 
We design an enhancer that efficiently enhances the Tetris-like irregular regions composed of important macroblocks (\S \ref{sec:EnhancementExecutor}). 
It prioritizes and selects the Top-K MBs of all video streams in order of their importance scores (K is estimated by \S \ref{sec:ComponentResourceManager}).
Then, considering the unique characteristic of the time cost of enhancement (\S \ref{sec:OurChoice}) and the sparse distribution of important MBs (\S \ref{sec:RegionSelector}), the region-aware enhancer can be formulated as a two-dimensional bin-packing problem that minimizes the input size of the enhancement model to optimize the throughput.  
We propose a greedy algorithm that fast stitches the irregular regions into dense tensors before forwarding them into the enhancement model.

\emph{Profile-based execution planning.} 
To balance the limited resources among components on the edge device, we propose an execution planer (\S \ref{sec:ComponentResourceManager}), that profiles the capacity of the given edge device, and allocates resources among components by determining the parameters (\eg, batch size of the input) for each component (\eg, decoder, MB-based region importance prediction, region-aware enhancer, and analytical models) to maximize the end-to-end throughput.



We summarize our key contributions as follows:

\scalebox{0.8}{$\bullet$} To the best of our knowledge, we are the first to identify the bottleneck of content-enhanced video analytics and propose a new idea: region-based content enhancement.

\scalebox{0.8}{$\bullet$} We prototype \name that enables region-based content enhancement video analytics at the edge. 
It involves three components, MB-level region importance prediction, region-aware enhancement, and profile-based execution planning.

\scalebox{0.8}{$\bullet$} We implement \name and conduct evaluations for two popular analytical tasks with real-world videos on five heterogeneous devices. 
Experimental results show that \name improves 10-19\% accuracy and achieves 2-3$\times$ throughput compared to the state-of-the-art methods. 
Our code is now available at \url{https://github.com/mi150/RegenHance}.


\textbf{This work does not raise any ethical issues.}

\section{Motivation and Challenges}
This section explores three questions: 
(1) How well are selective enhancement techniques in video analytics (\S \ref{sec:bottleneck})?
(2) How much potential improvement can region-based content enhancement achieve (\S \ref{sec:OurChoice})?
(3) What challenges must be tackled (\S \ref{sec:challenges})?

\subsection{Background}
\textbf{Typical video analytics platforms}~\cite{Rocket, GoogleAI, Dai2022BigDL}, to support real-time applications, 
provide services on edge servers close to cameras that allow users to register custom video analytics jobs. 
To register a job, users specify video sources, analytical tasks, performance targets, and optionally upload their analytical models~\cite{agarwal2023boggart, poddar2020visor, cangialosi2022privid}.
The camera captures and encodes multiple images into a video stream (\eg, 30 images into a 1-second video chunk set up in prior studies \cite{du2020server, li2020reducto, zhang2018awstream}), and then transmits to the edge for analytics.
Following the same scope, this paper also concentrates on optimizing the analytics platforms on edge servers.

\noindent
\textbf{Content-enhanced} video analytics~\cite{AccDecoder, kim2020neural, dasari2020streaming, yi2020eagleeye, wang2019bridging, yeo2020nemo, yeo2022neuroscaler, yeo2018neural, yi2020supremo, lu2022turbo} is a mainstream data optimization method, that utilizes DNN models to enhance low-resolution frames to high-definition ones. 
Pre-trained DNNs learn a pixel generator mapping neighboring pixels' values to the generated one from training data, then deployed in runtime systems analyzing real-world videos for accuracy improvement and bandwidth saving.


\noindent
\textbf{Selective enhancement} is the state-of-the-art method to improve the throughput of content-enhanced Internet video streaming~\cite{zhang2017fast, yeo2020nemo, yeo2022neuroscaler}.
In this method, a few sampled frames (called anchors) are enhanced by the super-resolution model, while the remaining frames are quickly up-scaled by reusing enhanced ones via codec information.
Such reuse causes video quality loss, due to the rate-distortion problem in codec \cite{sullivan1998rate}, accumulated across consecutive non-enhanced frames, and thus the sampled frames need to be selected carefully. 

\subsection{Limitation of Frame-based Enhancement}\label{sec:bottleneck}

\begin{figure*}[t]
\centering
  \begin{minipage}[b]{0.24\linewidth}
  \centering
    \includegraphics[width=1.0\linewidth]{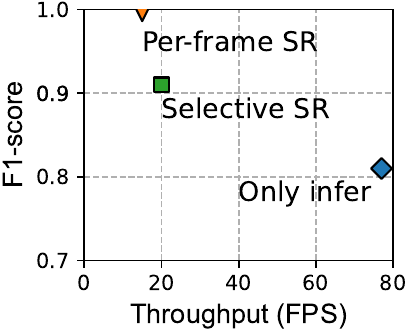}
    \caption{Performance of the frame-based methods. The state-of-the-art selective SR provides poor E2E throughput.}
      \label{fig:AccTPT}
  \end{minipage}
  \hfill
  \begin{minipage}[b]{0.24\linewidth}
  \centering
  \includegraphics[width=1.\linewidth]{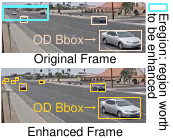}
    \caption{An example eregion worth to be enhanced for object detection (OD) bounding with a rectangle box.}
    \label{fig:IdealRegion}
    \end{minipage}
  \hfill
  \begin{minipage}[b]{0.24\linewidth}
  \centering
  \includegraphics[width=1.0\linewidth]{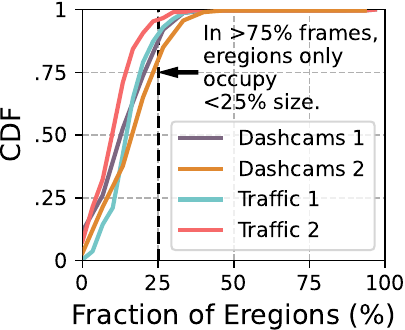}
    \caption{Distribution of eregions in various videos. In large amounts of frames, eregions occupy only a small portion.}
    \label{fig:DistributeinoOfIdealRegion}
    \end{minipage}
\hfill
  \begin{minipage}[b]{0.24\linewidth}
  \centering
    \includegraphics[width=1\linewidth]{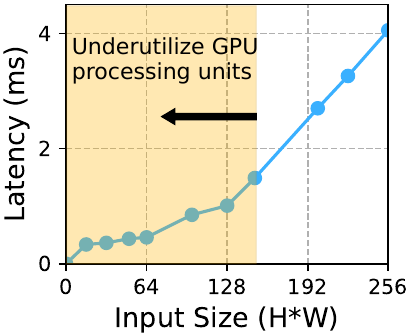}
    \caption{Latency of enhancement. The same H*W (\eg, 64*64) input, no matter pixel values, yields the same latency.}
      \label{fig:SRtime}
  \end{minipage}
  \vspace{-1em}
\end{figure*}
  
\noindent
\textbf{Motivational study.} 
To measure the performance of selective enhancement, we benchmark three analytics methods. 
Taking object detection as an example,
1) \emph{Only infer} method purely detects objects in each frame, 
2) \emph{Per-frame SR} method applies super-resolution on each frame and then detects objects on them (as the ground truth).
3) \emph{Selective SR}~\cite{yeo2022neuroscaler} applies super-resolution on a set of selected frames and reuses on other ones, then detects objects on all frames; this method will select enough frames until meeting preset target accuracy (\eg, 90\%).
For fairness, all methods run on an edge device equipped with an NVIDIA T4 GPU~\cite{T4} that up-scales a $640\times360$ video to a $1920\times1080$ version via EDSR model~\cite{Lim2017EDSR} and detects objects with YOLO~\cite{glenn2021YOLOV5}.
Both models compile with TensorRT~\cite{TensorRT}.


\noindent
\textbf{Result: selective enhancement is too heavy for video analytics.}
As shown in Fig. \ref{fig:AccTPT}, per-frame SR improves >10\% accuracy but reduces >76\% end-to-end throughput than only infer.
Normalized with per-frame SR, selective SR indeed improves throughput from 15 to 20 (>33\%) fps, 
but still far from the throughput of only infer and results in an undeniable accuracy drop.
This is because, compared to human perception, analytical models are more sensitive to blur and distortion caused by accumulated loss of reuse. 
Small changes in several pixel values may flip the analytics result inversely \cite{brown2017adversarial}.
So when presetting the accuracy target as 90\%, selective SR methods choose 27-61 anchor frames in each 120-frame chunk on average. 
Such 24-51\% fraction
is much higher than that select and enhance only 2-13\% frames 
in video applications for human vision.




\subsection{Potential Improvement of Region-based Content Enhancement}\label{sec:OurChoice}
Above \emph{frame-based} optimizations fail to offer efficient analytics because they evenly enhance every pixel in each frame.
We argue that this is unnecessary and causes a significant waste of computational resources. 

Our first observation is that \emph{the regions that are worth being enhanced in each frame occupy only a small portion.}
In this paper, we name the region that provides higher analytical accuracy after enhancement, \emph{Eregion}.
As visualized in Fig. \ref{fig:IdealRegion}, the frame after enhancement (bottom) can provide more detected objects, and the eregion (blue rectangle) is a simple region (discussed more in Appx. \ref{sec:AdditionalResults1}).
Fig. \ref{fig:DistributeinoOfIdealRegion} analyzes the distribution of eregions in experimental videos for Fig. \ref{fig:AccTPT}.
In >75\% of frames over videos recording various scenarios, eregions for object detection task only occupy 10-25\% of the spatial area of a frame; while for semantic segmentation, as illustrated in Appx. \ref{sec:AdditionalResults1}, only 10-15\% area in 70\% frames is eregions.
With careful design, eregions could minimize the enhanced area while offering comparable accuracy to per-frame SR. 
So ideally, only eregions should be enhanced in video analytics.

The second observation is that \emph{the time cost of enhancement DNNs is positively correlated with input size.}
As shown in Fig. \ref{fig:SRtime}, along with the enlarging input size, the enhancement latency initially experiences a gradual rise and then scales proportionally with the input size after making full use of processing units.
This unique characteristic originates from the nature of enhancement models that generate new pixels from neighboring ones with the learned mapping.
Thus, we must minimize the size to decrease enhancement latency.

\noindent
\textbf{Our idea.} 
With the above two observations in mind, we propose \emph{region-based content enhancement} that aims to only enhance the eregions to maximize the throughput of content enhancement while providing comparable accuracy to per-frame enhancement. 
The question next is how to fully explore the potential room for improvement.

\subsection{Challenges to Achieve Region-based Content Enhancement}\label{sec:challenges}

Building an effective system based on region-based content enhancement involves three challenges. 

\begin{figure}[t]
\centering
     \centering
    \includegraphics[width=0.235\textwidth]{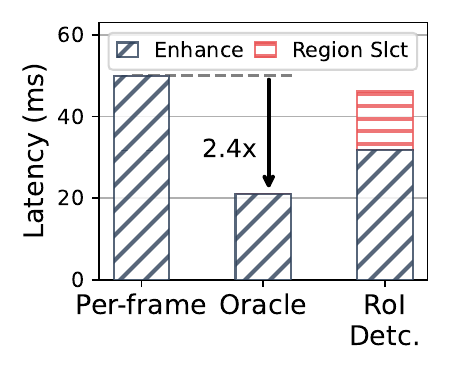}
    \caption{Region-based enhancement saves remarkable (2.4$\times$) latency, but prior region selection methods themselves (\eg, RoI detection~\cite{du2020server}) produces too high computing cost.}
    \label{fig:LatBreakdownOfPreviousMeth}
    \vspace{-1em}
\end{figure}

\noindent
\textbf{C1: How to fast and accurately identify eregions on original frames?}\label{sec:IllconsideredRegionEnhancement}
Enhancing only the eregions, as shown in Fig. \ref{fig:LatBreakdownOfPreviousMeth}, does save significant ($2.4\times$) time cost. 
Unfortunately, such an oracle cannot be reached as eregions are calculated from the already-enhanced frames.
During real-time analytics, only the original frames are accessible, 
thus we have to predict eregions 
on the original frames. 
Naively using a DNN-based method to identify regions, \eg, DDS~\cite{du2020server} identifies the Region of Interest (RoI) with a Region Proposal Network (RPN), can not meet our requirements.
The imprecise identification spends too much time on the enhancement of unimportant regions in Fig. \ref{fig:LatBreakdownOfPreviousMeth}, and the high computing cost of the selection methods themselves (\eg, the RPN), in turn, harms the possible time saving of region-based enhancement. 


\noindent
\textbf{C2: How to enhance eregions in high throughput? }\label{sec:IllconsideredRegionEnhancementb}
High-throughput region-based enhancement is hardly achieved by existing frame-based enhancers~\cite{yeo2018neural, yeo2022neuroscaler, yeo2020nemo, wang2022enabling, yi2020supremo, yi2020eagleeye}.
Three new contradictions that frame-based methods have never encountered obstruct region-based enhancement.
1) Eregions are often in irregular size and sparsely distributed in each frame, but the enhancement models (DNNs) only accept rectangle (matrix) input.
2) One natural option, as DDS \cite{du2020server} used, is setting non-eregion pixels to black (zero value in the matrix); and yet it leads to the same latency as enhancing the original frame due to the pixel-value-agnostic characteristic of enhancement latency as demonstrated in Fig. \ref{fig:SRtime}.
3) Batch execution of enhancement DNNs benefits higher throughput \cite{shen2019nexus}, but requires every input matrix in the same size. 


\noindent


\noindent
\textbf{C3: How to allocate resources among components?}\label{sec:InefficientResourceManagement}
To optimize the overall performance (accuracy and throughput), particularly for all video streams, 1) eregions must be carefully allocated across streams, and 2) computational resources must be balanced among computing components.
However, this is challenging due to the heterogeneity of regions and components.
First, the accuracy gains and enhancement overhead of regions are heterogeneous.
Second, all runtime components on the edge server, including decoding, eregion identification, enhancement, and analytical inference, compete for computational resources, and their throughputs are heterogeneous. 
Using previous schedulers \cite{K8Edge, yeo2022neuroscaler} causes 
noticeable degradation in accuracy and throughput.


To demonstrate this, we benchmark a strawman scheduler that 
1) parallelizes per-stream decoding on multiple CPU threads, 
2) forwards decoded frames to GPU in a round-robin manner, 3) applies \name's region enhancement strategy but pipelines computing components at batch size of four. 
Two streams that contain different sizes of eregions are delivered to an edge server equipped with an NVIDIA T4 GPU \cite{T4}. 
We measure the accuracy gain per stream and the resource utilization of computing units.

\begin{figure}[t]
\centering
\setlength{\abovecaptionskip}{0pt}
\hspace{-1em}
    \subfigure[Round robin across streams] {
    \centering     
     \label{fig:AccGain}     
    {\includegraphics[width=0.225\textwidth]{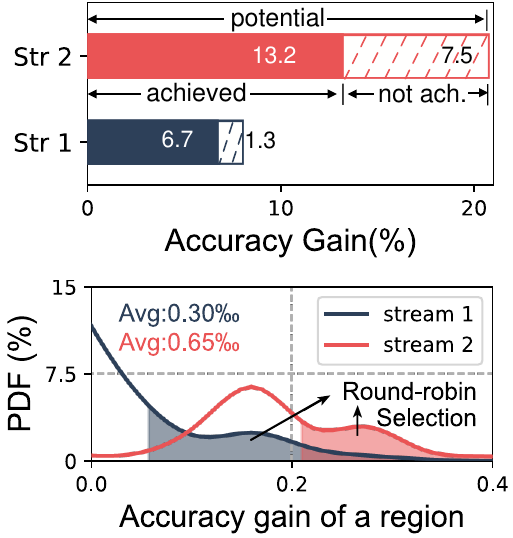}}
    }
    \subfigure[Sequential exec. of components] {
    \label{fig:EvenAllocationAmongComponents}
     \centering
     \raisebox{0.05cm}
    {\includegraphics[width=0.235\textwidth]{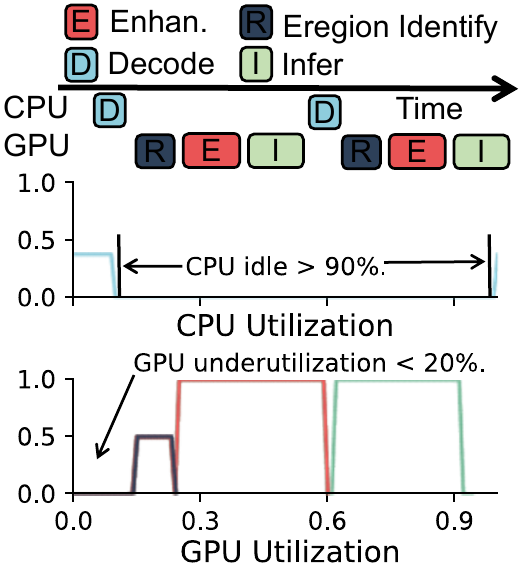}}
    }
    \caption{Limitation of region-agnostic resource management.}
    \label{fig:InefficientResourceManage}
    \vspace{-1em}
\end{figure}

This strawman region-agnostic scheduler fails to achieve high accuracy gain and throughput.
Fig. \ref{fig:AccGain} (top) shows the potential (the accuracy of per-frame SR minus only infer), achieved (solid part), and not achieved (dotted part) accuracy gain of two streams.
There is a remarkable gain (7.5\%) not achieved in Stream 2
as the round-robin manner results in an even chance for enhancement across streams.
The distributions of the region's accuracy gain in each stream are highly heterogeneous as the two curves shown in Fig. \ref{fig:AccGain} (bottom). 
Suppose a scheduler enables Stream 2 to enhance more regions that provide higher gain (namely, in Fig. \ref{fig:AccGain} (bottom), decrease some grey area under Stream 1 to fill the blank under Stream 2 into red), it can improve the overall accuracy.
On the other hand, 
inappropriate execution plans underutilize the hardware, leading to low end-to-end throughput and, in turn, hindering higher accuracy improvement.
As illustrated in Fig. \ref{fig:EvenAllocationAmongComponents}, the region-agnostic scheduler leaves >90\% CPU and >15\% GPU idle time.
The unsuitable batch size results in the Eregion Identification utilizing only <50\% computing resource GPU, hence blocking the subsequent enhancement and analytics.




\vspace{-1em}
\section{\name Design}
\begin{figure}[t]
\centering
\includegraphics[width=0.8\linewidth]{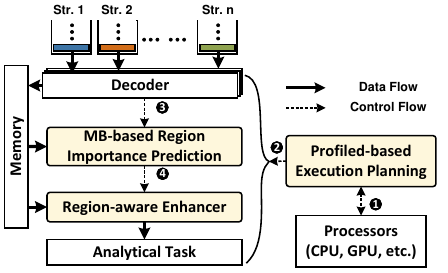}
    \caption{\name overview.}
    \label{fig:overview}
    \vspace{-1em}
\end{figure}

\name aims to enhance only the regions (Eregions) that best benefit analytical accuracy, so as to maximize the overall accuracy gain among all streams in the optimal end-to-end throughput on a given edge server.
To do so, it addresses the above challenges by three techniques: 1) a lightweight but precise algorithm to identify eregions on macroblock (MB) granularity, 2) a high-throughput enhancer that together considers the heterogeneous accuracy gain across streams and unique characteristics of enhancement, and 3) a holistic resource manager generating execution plan for all components on given edge servers.
Fig. \ref{fig:overview} depicts the design of \name.
We implement the techniques with three new components: MB-based region Importance Prediction, Region-aware Enhancement, and Profile-based Execution Planning.



\subsection{Overview Workflow}
\name takes the original ingest videos and outputs high-resolution frames for downstream analytical tasks.
During the offline phase, Profile-based Execution Planning \ding{182} profiles the budget of processors on the given edge server with the registered analytical tasks and optional uploaded corresponding models by users; next, \ding{183} generates the execution plan, \ie, specifies every runtime component's execution hardware and their allocated resource meeting best pipelining and parallelization, to maximize the end-to-end throughput under the constraints of performance targets (\eg, accuracy and latency) that users specify (\S \ref{sec:ComponentResourceManager}).

During the online phase, the decoder decodes compressed video streams in parallel to RGB frames and stores them in memory. 
MB-based region importance prediction \ding{184} selects frames during decoding, loads them from memory, predicts their MBs importance, and appends importance into MBs' indexes (\S \ref{sec:RegionSelector}).
Region-aware enhancement \ding{185} aggregates and sorts all MBs over all streams based on their importance, constructs top N ones into regions, stitches them into dense tensors, and efficiently executes enhancement (\S \ref{sec:EnhancementExecutor}).
Lastly, forward the enhanced frames to analytical models for inference.

\subsection{MB-based Region Importance Prediction}\label{sec:RegionSelector}
In order to identify eregions quickly and accurately, we propose an MB-based region importance prediction.
It contains a lightweight predictor estimating MB importance in each frame (in spatial dimension) and an MB importance reuse algorithm among continuous frames (in temporal dimension).

\subsubsection{Spatial MB Importance Predictor in Each Frame.}
Eregions can be arbitrary shapes, we need to determine the granularity to construct these regions first.
One natural way is pixel granularity.
The pixel-grain method, of course, provides the most precise importance prediction but costs too much computing resources (as demonstrated in Fig. \ref{fig:LatBreakdownOfPreviousMeth}). 

Inspired by the video codec knowledge \cite{du2022AccMPEG, hwang2022cova}, we argue that setting macroblock (MB) as the elementary unit of eregions is both efficient and accurate.
Macroblock serves as the elementary unit for applying the quantization parameter (QP) to control the compression level of video quality.
For example, in H.264~\cite{H.264}, frames are divided into an array of 16$\times$16-pixel MBs, and each MB is assigned with different QPs to distribute more bits to regions demanding higher visual quality and fewer bits to less crucial regions.

After setting MB as a unit, our problem transformed to: 
Given a video frame $f$ containing a set of macroblocks ${MB}$, select partial MBs $MB_{s}$ to be enhanced such that the accuracy of downstream analytical task is maximized:
\begin{equation}\label{equ:formulation}\small
\setlength{\belowdisplayskip}{3pt}
\begin{aligned}
max\quad Acc(I(SR(MB_{s})+IN(\overline{MB}_{s})), I(SR(f))), \notag
\end{aligned}
\end{equation}
where $Acc(\cdot)$ is the accuracy of analytical task $I(\cdot)$ (\eg, F1-score of object detection), $SR(\cdot)$ and $IN(\cdot)$ are the super-resolution and the bi-linear interpolation with the same enlarge factor, $\overline{MB}_{s}$ indicates the unselected MBs. 

\begin{figure}[t]
\centering
\setlength{\abovecaptionskip}{0pt}
\hspace{-0.5em}
    \subfigure[Heat map of MB importance] {
    \centering     
     \label{fig:MacroblockScoring}     
    \includegraphics[width=0.235\textwidth]{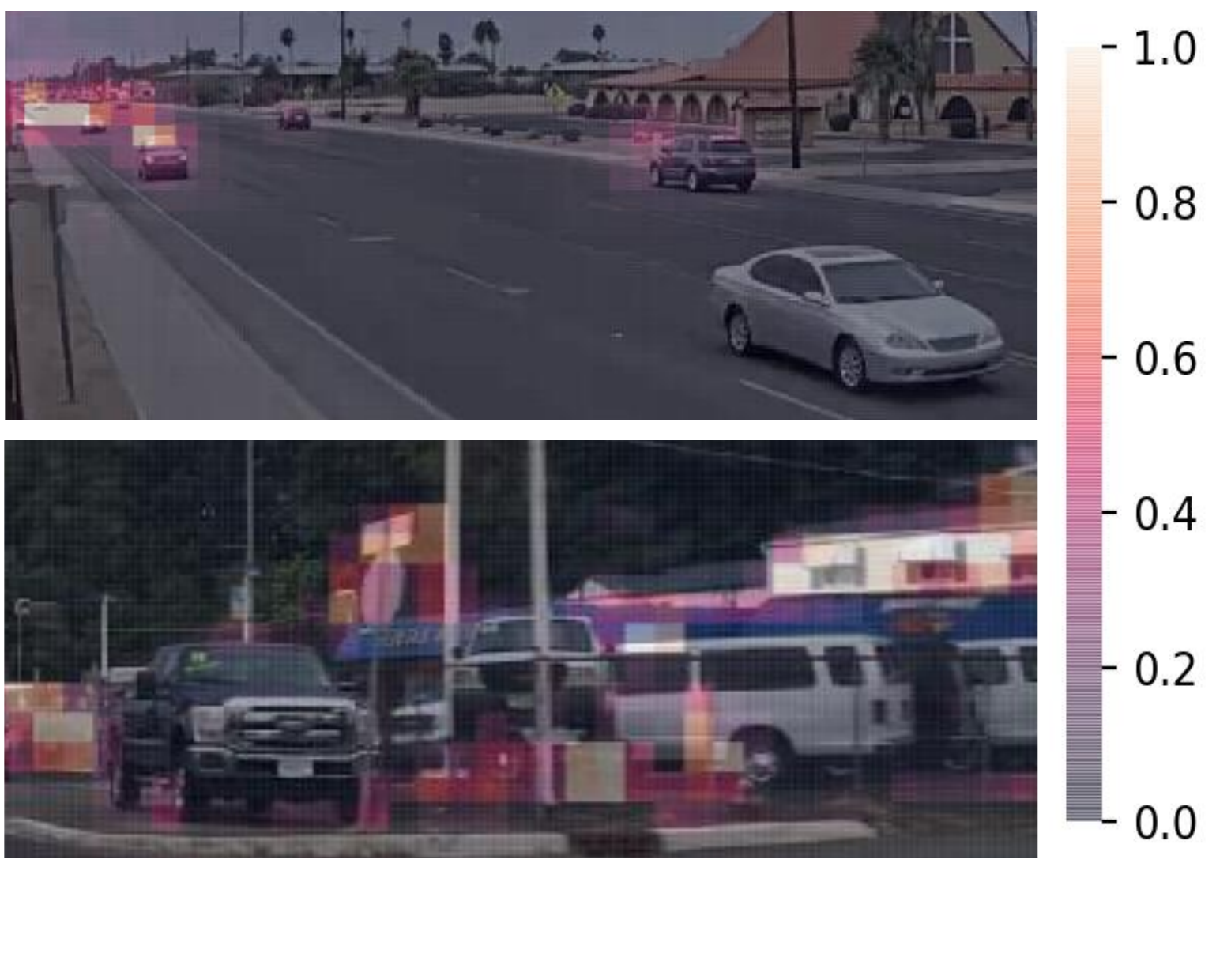}
    }
    \hspace{-1em}
    \subfigure[Model selection] {
    \label{fig:ModelSelection}
     \centering
    \includegraphics[width=0.235\textwidth]{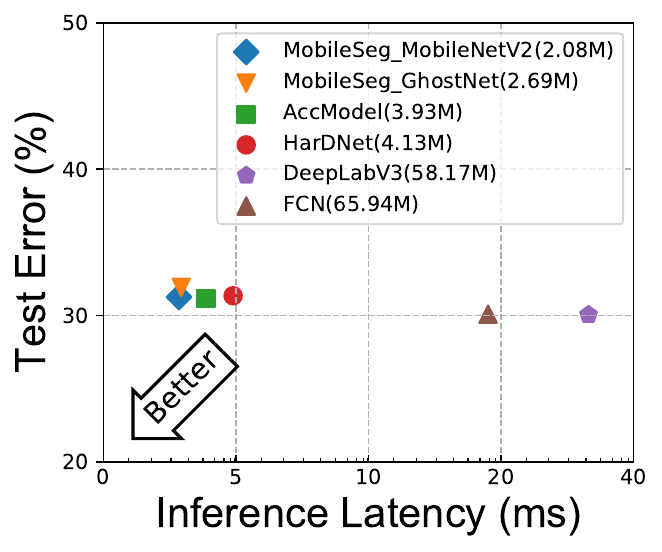}
    }
    \caption{Macroblock-level region importance prediction.}
    \label{fig:xxx}
    \vspace{-0.7em}
\end{figure}

To select beneficial MBs, we need to quantify their importance. 
Ideally, MBs after enhancement leading to larger variations in inference results and greater differences in pixel values are more important.
With this intuition, we utilize the following ``importance'' metric looking at the gradient of the accuracy with respect to changes in the pixels in each MB, and the magnitude of change in those pixel values due to enhancement.
\begin{equation}\label{equ:metric}\small
\begin{aligned}
\sum_{i\in MB}||\underbrace{\frac{ \partial Acc(I(IN(f)), I(SR(f))) }{ \partial IN(f)}\Big|_i}_{\text{\parbox{3cm}{accuracy's gradients at pixel $i$}}}||_1 \cdot ||\underbrace{(SR(f)_i-IN(f)_i)}_{\text{{pixel value distance at $i$}}}||_1, \notag
\end{aligned}
\end{equation}
where $i$ is a pixel within the certain MB and $||\cdot||_1$ is the L1-norm. 
This metric assigns an importance score to each MB. 
Fig. \ref{fig:MacroblockScoring} shows the heat map of MB importance on the same frame with Fig. \ref{fig:IdealRegion}.  
Their similarity demonstrates the MB importance 
is a good representation of eregions\footnote{Unlike the saliency map in computer vision that captures which pixel values have more influence on the DNN output, our loss function captures how the enhancement or non-enhancement of an MB (its content quality change) specifically changes the DNN inference accuracy.}.

It seems straightforward to calculate the MB importance according to the above importance metric. 
Unfortunately, it needs the frames already enhanced.
Such a chicken-egg paradox makes us predict the MB importance in the original frames with a learning-based method.
We construct the training set by enhancing all video frames and calculating the importance metrics on each of them with one forward and backward propagation of the final analytical model. 
This resulting importance value of each MB is one item in the ground truth matrix of each frame, Mask*.


Next, we consider this problem a segmentation task and draw inspiration from their model designs.
Our key observation is that the MB importance prediction resembles the problem of image segmentation.
Image segmentation aims to semantically segment an image by assigning each pixel one predefined label, while MB importance prediction gives each MB an importance score. 
With this in mind, the MB importance prediction can be approximated as an image segmentation problem by boiling the importance value down to multiple importance levels. 
Suppose setting ten levels in this paper, MB importance prediction targets to assign each MB an importance level, like classification in image segmentation.
Appx. \ref{sec:Derivation} demonstrates such approximation yields good performance.
This observation enables us to harness various techniques tailored for MB importance prediction, including learning-based semantic segmentation.




Aiming to predict MB importance precisely at high throughput, we train an ultra-lightweight segmentation model with the above importance metric.
We retrained six models using the cross-entropy loss with piecewise Mask* (as importance level) to support MB importance prediction.
Six models are an ultra-lightweight model 
MobileSeg~\cite{tang2023pp} with two backbones, two lightweight models 
AccModel~\cite{du2022AccMPEG} and  HarDNet~\cite{chao2019hardnet}, 
and two heavyweight models 
FCN~\cite{long2015fully} and DeepLabV3~\cite{chen2017rethinking}. 
As shown in Fig. \ref{fig:ModelSelection}, ultra-lightweight models provide almost the same accuracy as heavyweight ones while offering 4-18$\times$ throughput.
This can be attributed to the significantly relaxed complexity of the MB-grained segmentation compared to traditional image segmentation. 
Predefined MB size in video codec supports this point.
The 16$\times$16-MB H.264~\cite{H.264} codec makes 1920$\times$1080 labels that output in traditional image segmentation models decrease to 120$\times$68 ones in MB-grained.
As the best performance of MobileSeg, we select it as our MB importance predictor.
At the offline phase, RegenHance fine-tunes the predictor with the Mask* generated by user-uploaded analytical models.




\begin{figure}[t]
  \centering
  \setlength{\abovecaptionskip}{0pt}
\hspace{-1em}
    \subfigure[Correlation ratio] {
    \centering     
     \label{fig:CorrelateRatio}     
    \includegraphics[width=0.46\linewidth]{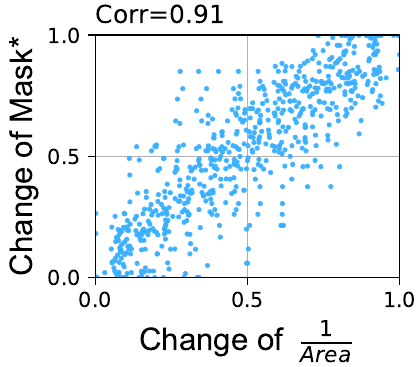}
    }
    \subfigure[Frame selection with CDF] {
    \label{fig:CDFSelect}
     \centering
    \includegraphics[width=0.47\linewidth]{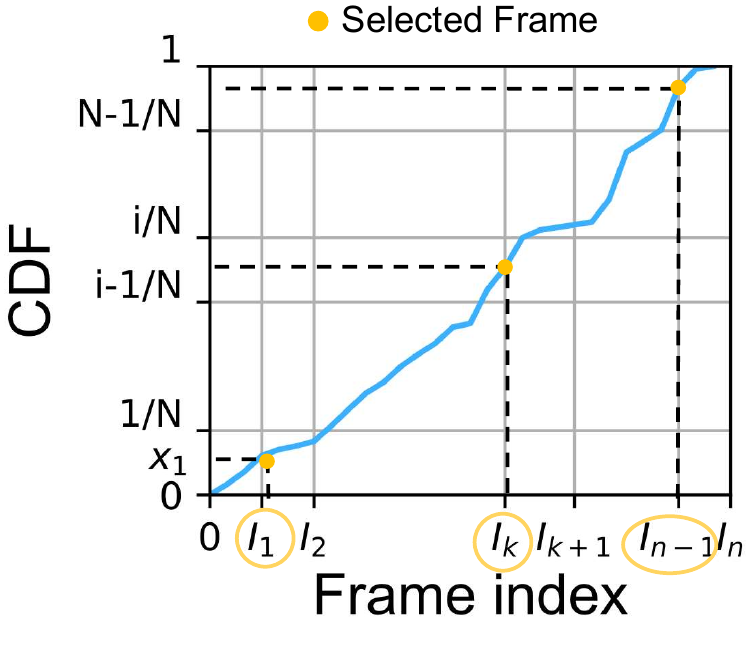}
    }
    \caption{Temporal MB importance reuse across frames.}
    \label{fig:FrameSelection}
    \vspace{-1.3em}
\end{figure}


\subsubsection{Temporal MB Importance Reuse among Continuous Frames.}\label{sec:TemporalSelection}
Reusing DNN outputs on similar frames is popular \cite{du2022AccMPEG, AccDecoder, xu2018deepcache} as discussed in \S \ref{sec:RelatedWorks}.  
Reusing content of enhanced frames in region-based enhancement like selective SR will cause dramatic accuracy loss (demonstrated in \S \ref{sec:bottleneck}), but the MB importance value is reusable.
Hence, RegenHance predicts the MB importance of a set of frames and reuses their outputs on other frames, to offer the best approximation of per-frame MB importance prediction. 

Our key choice is using an ultra-lightweight operator to represent the MB importance change then only predicts MB importance of the frames with large changes. 
We compared many lightweight features and proposed $\frac{1}{Area}$ operator (details in Appx. \ref{sec:AdditionalResults2}).
Area operator captures the large blocks in images~\cite{hu2011survey, li2020reducto}, whereas $\frac{1}{Area}$ captures the change of small objects just as the important MBs in Fig. \ref{fig:MacroblockScoring} needed. 
Statistical analysis in Fig. \ref{fig:CorrelateRatio} shows it is a nice representation to estimate the change of Mask* (with $0.91$ correlation).

With the $\frac{1}{Area}$ operator, denoted by $\Phi$, \name selects frames to be predicted by the cumulative distribution function (CDF) of $\Delta\Phi$ between continuous frames in each chunk.
It first accumulates the feature change with the following function during decoding a $n$-frame chunk.
\begin{equation}\small
\setlength{\abovedisplayskip}{5pt}
\setlength{\belowdisplayskip}{5pt}
\begin{aligned}
\mathbf{S}=Norm(\Delta\Phi(Res_{Y_{1}}),\cdots,\Delta\Phi(Res_{Y_{n-1}})), \notag
\end{aligned}
\end{equation}
where \small$Res_{Y_i}$\normalsize is Y-channel of each frame's residual\footnote{Residual is data existing in video codec representing the difference data between the original and compressed video frames.}, \small$\Delta\Phi(Res_{Y_i})=\Phi(Res_{Y_{i+1}})-\Phi(Res_{Y_i})$\normalsize, 
\small$Norm(\cdot)$\normalsize\ is L1-normalization.
Then, it selects $N$ frames based on the CDF \small$\mathbf{M}$\normalsize, where \small$\sum_i M_i=1$\normalsize, calculated from \small$\mathbf{S}$\normalsize.
As illustrated in Fig. \ref{fig:CDFSelect}, the y-axis is divided into $N$ even intervals; and in each interval, it selects a value, \eg, $x_1$, and then its corresponding frame index, \eg, $I_1$, on the x-axis is a selected frame.
The other frames in each interval reuse the predicted result of this frame.
In multiple streams, the number of selected frames for given stream $j$ is allocated by the ratio $\frac{\sum_i\Delta\Phi_i,j}{\sum_j\sum_i\Delta\Phi_{i,j}}$, and its total number is determined by the profile-based execution planning (\S \ref{sec:ComponentResourceManager}).
Fig \ref{fig:RuntimeRegenHance} depicts the MB-based importance prediction first selects frames and then predicts their MB importance.


\subsubsection{Discussion.}
\emph{Generality of importance metric.}
In this paper, each analytical task must retrain a specific MobileSeg for importance prediction, as the metric equation relies on the downstream models.
We omit a general importance metric design in this paper because 1) the current specific metric offers good results (as shown in Fig. \ref{fig:MacroblockScoring}) and 2) the fine-tuning time only costs 4 minutes on eight RTX3090 GPUs.
We will explore a general metric in future work.





\subsection{Region-aware Enhancement}\label{sec:EnhancementExecutor}
%
%
To achieve optimal overall accuracy over all streams and end-to-end throughput, 
we propose a region-aware enhancement consisting of 
a cross-stream MB selection 
and a region-aware bin packing algorithm. 




\subsubsection{Cross-stream MB Selection.}
\begin{figure}[t]
\centering
\hspace{-1em}
  \centering
\includegraphics[width=1.25\linewidth]{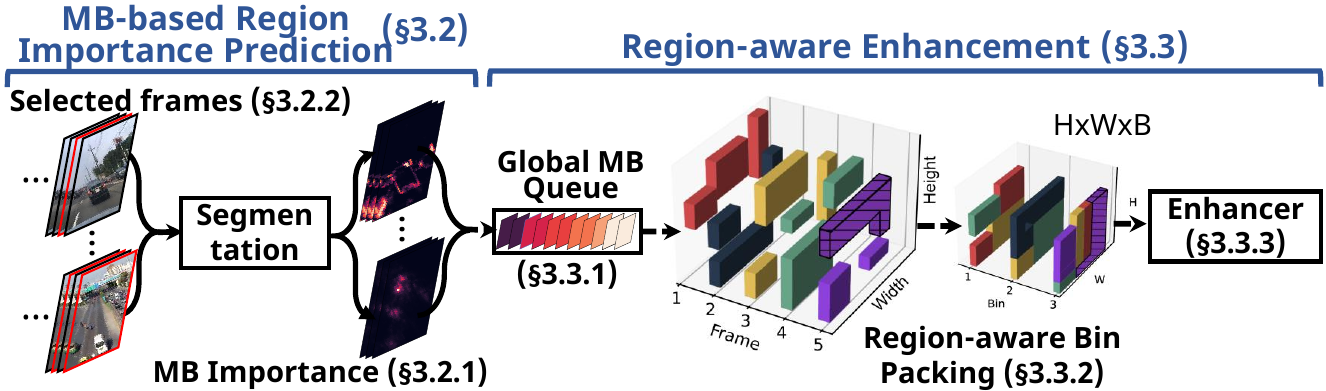}
    \caption{Runtime components of \name.}
\label{fig:RuntimeRegenHance}
  \end{figure}

To maximize the overall accuracy improvement, 
\name chooses the best MB st among all streams that offer the highest total accuracy gain. 
As shown in Fig. \ref{fig:RuntimeRegenHance}, it constructs a global queue that aggregates and sorts MBs from all streams in order of the importance (level) in the MB index {\small $\{stream_{id}, frame_{id}, loc_x, loc_y, importance\}$}, where {\small $loc_x, loc_y$} indicates coordination of the MB in the frame. 

Next, it selects the top $N$ MBs 
and delivers their indexes to the region-aware bin packing algorithm for further processing. 
The number of selected MBs is estimated as follows,
\begin{equation}\small
\setlength{\abovedisplayskip}{5pt}
\setlength{\belowdisplayskip}{5pt}
\begin{aligned}
    \max_{N}\ MB_{size}\cdot N \leq H\cdot W\cdot B, \notag
\end{aligned}
\end{equation}
where $MB_{size}$ is the size of a MB, \eg, 16$\times$16 in H.264. 
$H, W, B$ are the optimal height, width, and batch size for the enhancement model preset in the execution plan (\S \ref{sec:ComponentResourceManager}). 

\begin{algorithm}[t]\footnotesize
\caption{Region-aware Bin Packing}
\label{alg:binpacking}
\begin{algorithmic}[1]
\algnotext{EndIf}
\algnotext{EndFor}
\algnotext{EndFunction}
\Require Indexes of MBs, $B$ Bins of $H\times W$ size
\Ensure Packing plan of MBs
\Function{Packing}{MBs, Bins}  
\State freeareas = Bins \Comment{Initialize the list of free area as Bins}
\State regions = \Call{RegionProps}{MBs} 
\State boxes = \Call{Bound}{regions}
\State boxes = \Call{Partition}{boxes} \Comment{Partition big boxes to small ones}
\State \Call{Sort}{boxes, reverse = True, order = $\frac{\sum_{MB\in box}{MB.importance}}{|\{MB\mid MB\in box\}|}$} 
\For {box \textbf{in} boxes}
    \For {farea \textbf{in} freeareas}
        \If{\Call{RotatePacking}{box, farea}} 
        \State \Call{Update}{farea, freeareas}; \textbf{break}
        \EndIf
    \EndFor
    \State boxes.\Call{Pop}{box}
\EndFor
\EndFunction
\Function{RotatePacking}{box, farea}
\If{farea.w$\geq$box.w \textbf{and} farea.h$\geq$box.h} \textbf{return} True
\ElsIf{farea.w$\geq$box.h \textbf{and} farea.h$\geq$box.w} \textbf{return} True
\Else \textbf{ return} False
\EndIf
\EndFunction
\Function{Update}{farea, freeareas, box} 
    \State box.\Call{Append}{loc} \Comment{Append the placed location of box in Bins}
    \State ifareas = \Call{InnerFree}{farea, box}   
    \Comment{\parbox[t]{.45\linewidth}{Find rest free areas in box after packing farea (more in Appx. \ref{sec:FunctionsinAlg})}}
    \State freeareas = freeareas $\cup$ ifareas $\setminus$ \{farea\}
\EndFunction
\end{algorithmic}
\end{algorithm}
\setlength{\textfloatsep}{8pt}

\subsubsection{Region-aware Bin Packing.}\label{sec:BinPacking}
Considering the sparse distribution of selected MBs, the unique characteristics of enhancement models (requires rectangle input and latency is pixel-value-agnostic proportional to the input size), and the helpful but complex batch execution, we propose a region-aware bin packing algorithm to construct selected MBs into irregular regions and stitch them into dense tensors.



The problem is formulated as a \emph{two-dimensional bin-packing problem} \cite{berkey1987two} that packs the maximum number of selected MBs into given bins. 
The input are
MB indexes and the 
number of bins $B$ and their size $H\times W$, the output is the packing plan of MBs. 
We process MB indexes, not real images, to avoid frequent memory I/O.
It is known to be NP-hard.
Prior methods \cite{cui2022dvabatch, shen2019nexus, cui2021enable} batch the standard rectangle DNN inputs and thus can not handle irregular regions.
To motivate our design, we first analyze two strawmen: 1) MB packing: directly packing selected MBs after expanding three pixels in each direction\footnote{Such expansion can avoid the MB/region boundaries causing too many jagged edges and blocky artifacts \cite{du2022AccMPEG} when pasting enhanced content back to the bi-linear-interpolated frames. For more results refer to Appendix \ref{sec:AdditionalResults3}.} into bins, and 2) Irregular region packing: packing the (irregular) regions consisting of connected and selected MBs after extension into bins \cite{lopez2013effective}. 
MB packing packs too many unimportant or repeated pixels due to extension, thus under-utilizing given bins compared to the irregular manner; 
whereas the high complexity of existing irregular region packing algorithms causes more than one order of magnitude of time cost than MB packing (more in Appx. \ref{sec:AdditionalResults4}).



  \begin{figure}[t]
  \centering
  \hspace{5em}
\includegraphics[width=0.75\linewidth]{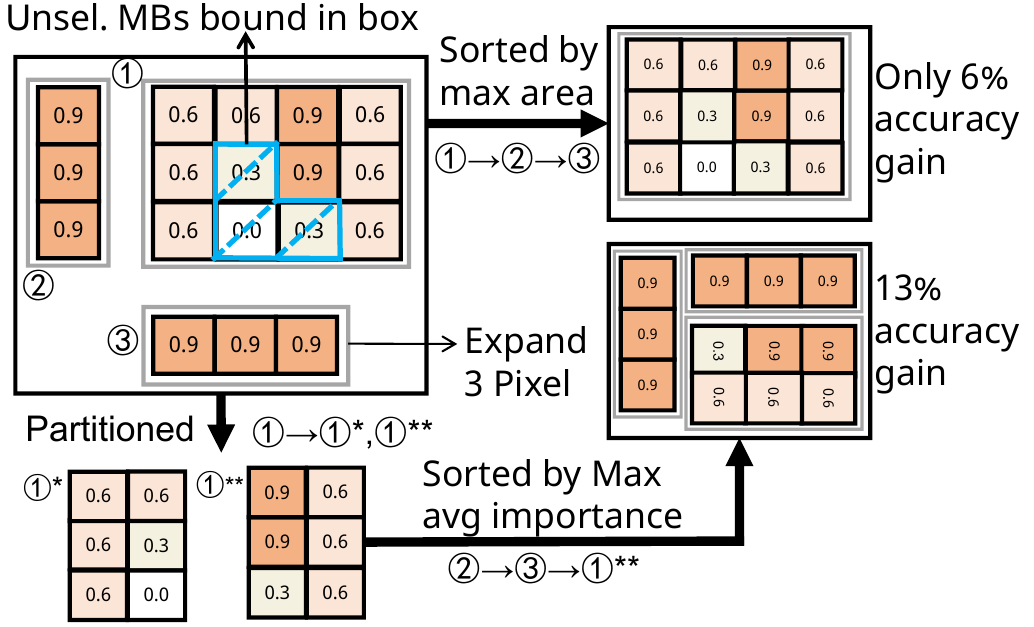}
    \caption{Large region is not always desirable.  Classic large-item-first policy packs the MBs that are unselected but bounded in the large box into the bin (upper half), leading to a 7\% accuracy drop compared to our importance-first policy (lower half).}
    \label{fig:Puzzle}
    \end{figure}

To strike a better balance between bin utilization and algorithm efficiency, we propose region-aware bin packing in Alg. \ref{alg:binpacking}. 
Its key design choices are: 1) bounding the irregular regions in rectangle boxes for efficient packing plan search (line \#3-5); 
2) cutting large boxes into small ones and sorting boxes in order of importance density, \ie, the average importance of all MBs in it, 
for high bin utilization (line \#6).
Fig. \ref{fig:Puzzle} exemplifies that this prioritization leads to a higher ($13\%$) accuracy gain, compared to $6\%$ of traditional large-item-first (max-area-first) policies~\cite{puzzle} (more results refer to Fig. \ref{fig:area_score}).
Region-aware bin packing first constructs regions by calculating the connected components 
of selected MB in line \#3, and bounds each region in a rectangular box with extension (line \#4), \eg, \ding{172}\ding{173}\ding{174} in Fig. \ref{fig:Puzzle}. 
Next, it partitions the boxes larger than the preset size (line \#5) in case of importing too many unimportant MBs, \ie, box \ding{172} cut to \ding{172}* and \ding{172}**.
Then, line \#6 sorts boxes according to the priority we propose, \ie, not (\ding{172},\ding{173},\ding{174}) of max-area-first policy but (\ding{173},\ding{174},\ding{172}**). 
Lastly, it iteratively packs them into bins.
In each iteration (line \#7-11), it
rotates and packs a box into a free area in a bin, 
updates the box's placing location and the list of free areas, and then deletes the box from the boxes.
For example, the ``L'' region in Fig. \ref{fig:RuntimeRegenHance} and \ding{172}** in  Fig. \ref{fig:Puzzle} is rotated and packed into bins.



\subsubsection{Enhancement with Super-resolution Model.}

Based on the execution plan across various devices (\S \ref{sec:ExperimentalSetup}), except for Jetson AGX Orin equipped with unified memory, the other devices need to transfer real frames from main memory to GPU memory. 
To save time, we hide this transfer at the same time as the MB selection and bin packing procedure.
This idea works because, after processing the real frames by MB-based region importance prediction, as shown in Fig. \ref{fig:overview}, all modules solely deal with MB indexes before super-resolution.
Consequently, we stitch the real-content regions into tensors (bins) following the packing plan on the GPU, subsequently enhancing and stitching them back to bi-linear-interpolated non-regions for final analytics.

\subsection{Profile-based Execution Planning}\label{sec:ComponentResourceManager}
Given an edge server equipped with $R$ computational resource (\ie, $R$ processing time of processors under 100\% utilization) and the user's analytical jobs, profile-based executing planning aims to allocate the most appropriate resources to each component, maximizing the end-to-end throughput subject to performance targets. 
The problem is formulated as follows,
\begin{equation}\small
\begin{aligned}
\max\ \ T_{e2e},\ \ s.t. \sum R_{u}\leq R\ \ \forall u\in V(G).\notag\\
\end{aligned}
\end{equation}
where $G$ is the dataflow graph (DFG) of components; $u$ is the node in graph $G$ and $R_u$ indicates its allocated resource.

\begin{figure}[t]
\centering
\includegraphics[width=0.45\textwidth]{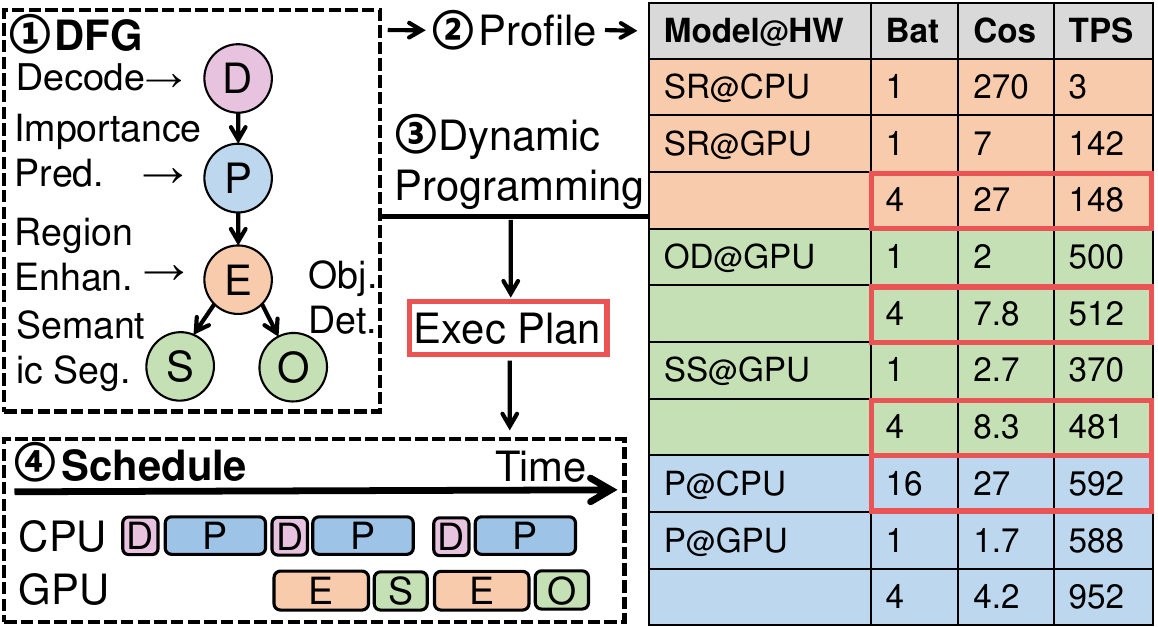}
    \caption{An example of execution planning process seven 30-fps streams s.t. <1s latency \& >0.9 accuracy performance target. The throughputs (TPS) of one model executing on one hardware (Model@HW) in the right table are profiled with batch size (Bat) and time cost (Cos). The red boxes highlight the execution plan of each component of the given dataflow graph (DFG).}
    \label{fig:manager}
\end{figure}

We present a profile-based execution planning, as the example of one CPU plus one GPU depicted in Fig. \ref{fig:manager}, that \ding{172} parses the DFG of user-specified analytical tasks, \ding{173} runs workload (\eg, one-minute user-specified streams) on all components, including user-uploaded models to profile their capacity on all accessible hardware, \ding{174} generates the execution plan that satisfies user-specified performance targets, and \ding{175} loads each component into corresponding hardware.
Our crucial choice is to assign each component's input tensor size (\eg, batch size) for resource allocation. 
Batching execution, \ie, grouping input matrices into one, is well known to achieve high processor utilization by DNNs; 
it also allows the inference engines to achieve various throughputs by adjusting different batch sizes~\cite{cui2022dvabatch, cui2019ebird, shen2019nexus}.

As the DFG of any job, naturally, is a directed acyclic graph (DAG), we can always leverage \emph{dynamic programming} to solve this optimization problem \cite{Vazirani2003Alg}.
Define $T_u(r)$ as the maximum throughput of components represented by $u$ and the subtree at $u$ within the resource budget $r$, which numerically equals the minimum node in each path. 
For non-leaf node $u$, the algorithm allocates a resource budget $r'$ for node $u$ and at most $r-r'$ for the subtree, and it then enumerates all $r\leq R$ to find the optimal allocations that assign optimal batch size $b$ for each node.
Formally, $\forall(u,v)\in E(G)$,
\begin{equation}\small
\begin{aligned}
T_u(r)=\max_{r:r\leq R}\left\{\min(\max_{b:c_u(b)\leq r'}\frac{b}{c_u(b)}, T_v(r-r'))\right\}. \notag
\end{aligned}
\end{equation}
where $c_u(b)$ is the resource cost of node $u$ at $b$ batches.
RegenHance allocates the least resources for analytical models that satisfy the user's latency target and then assigns other components' batch sizes with this equation.

In general, the optimal solution always converges to the allocation that won't be bottlenecked by any node; in other words, each node in the graph generates the same throughput.
Therefore, when users' registrations change frequently, the execution plan must be generated quickly.
To this end, we will explore learning-related methods like online deep reinforcement learning \cite{arulkumaran2017deep} and combinatorial-optimization-related methods like local search \cite{hoos2004stochastic} in future work.

\begin{figure*}[t]
    \centering
    \setlength{\abovecaptionskip}{0pt}
    \subfigure[4090] {
     \label{fig:4090}
    \begin{minipage}[b]{0.18\linewidth}
    \centering
    \includegraphics[width=1.1\linewidth]{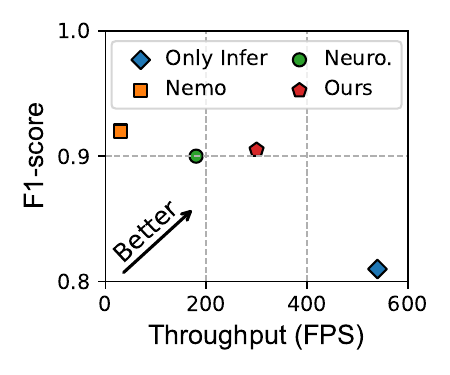}
    \end{minipage}
    }
    \hfill
    \subfigure[A100] {
    \begin{minipage}[b]{0.18\linewidth}
    \centering
    \includegraphics[width=1.1\linewidth]{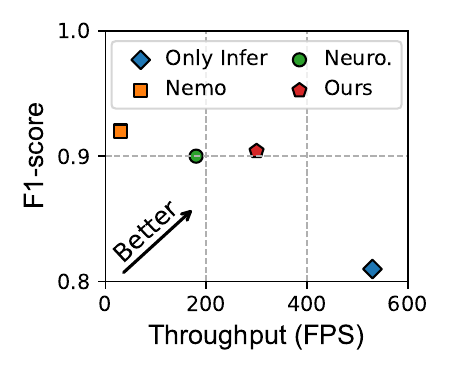}
    \end{minipage}
    }
    \hfill
    \subfigure[3090Ti] {
    \begin{minipage}[b]{0.18\linewidth}
    \centering
    \includegraphics[width=1.1\linewidth]{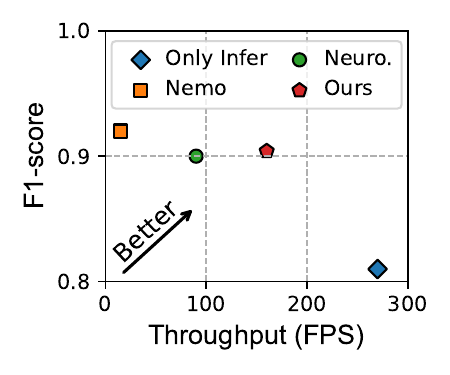}
    \end{minipage}
    }
    \hfill
    \subfigure[T4] {
    \begin{minipage}[b]{0.18\linewidth}
    \centering
    \includegraphics[width=1.1\linewidth]{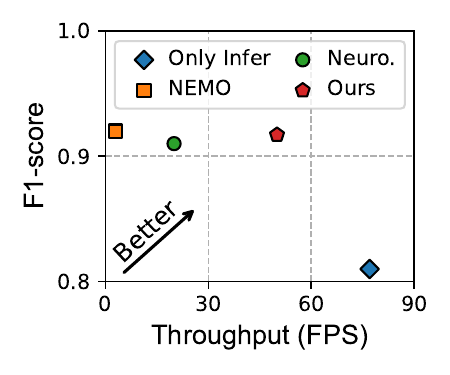}
    \end{minipage}
    }
    \hfill
    \subfigure[Jetson AGX Orin] {
    \begin{minipage}[b]{0.18\linewidth}
    \centering
    \includegraphics[width=1.1\linewidth]{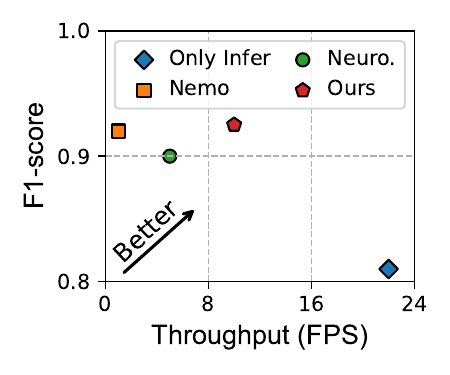}
    \end{minipage}
    }
    \caption{Accuracy and throughput comparison over various devices (Object detection).}
    \label{fig:odDevicecompare}
    \vspace{-1em}
\end{figure*}

\begin{figure*}[t]
    \centering
    \setlength{\abovecaptionskip}{0pt}
    \subfigure[4090] {
    \begin{minipage}[b]{0.18\linewidth}
    \centering
    \includegraphics[width=1.1\linewidth]{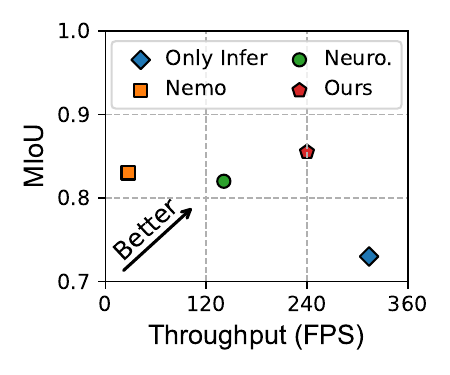}
    \end{minipage}
    }
    \hfill
    \subfigure[A100] {
    \begin{minipage}[b]{0.18\linewidth}
    \centering
    \includegraphics[width=1.1\linewidth]{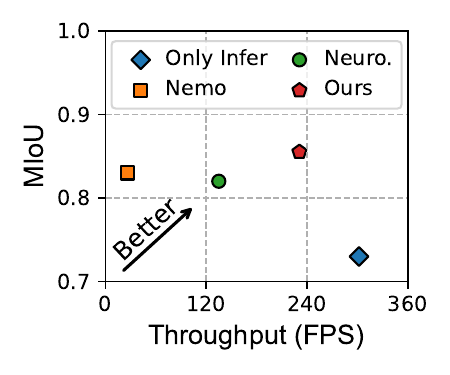}
    \end{minipage}
    }
    \hfill
    \subfigure[3090Ti] {
    \begin{minipage}[b]{0.18\linewidth}
    \centering
    \includegraphics[width=1.1\linewidth]{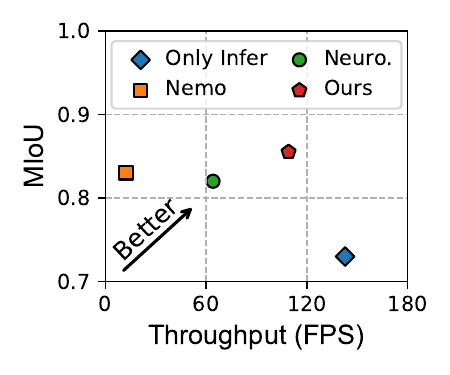}
    \end{minipage}
    }
    \hfill
    \subfigure[T4] {
    \begin{minipage}[b]{0.18\linewidth}
    \centering
    \includegraphics[width=1.1\linewidth]{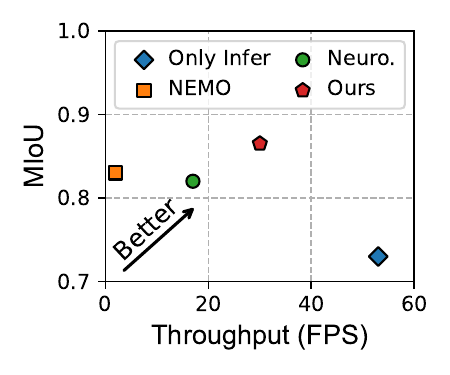}
    \end{minipage}
    }
    \hfill
    \subfigure[Jetson AGX Orin] {
    \begin{minipage}[b]{0.18\linewidth}
    \centering
    \includegraphics[width=1.1\linewidth]{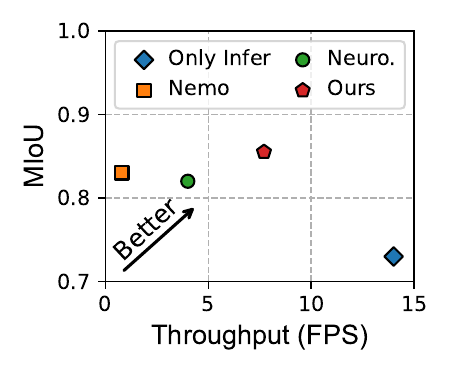}
    \end{minipage}
    }
    \caption{Accuracy and throughput comparison over various devices (Semantic
segmentation).}
    \label{fig:ssDevicecompare}
    \vspace{-1em}
\end{figure*}

\vspace{-0.5em}
\section{Evaluation}\label{sec:evalutaion}
\vspace{-0.5em}
We evaluate \name with two video analytics tasks on five heterogeneous devices. 
Our key findings are:

\scalebox{0.8}{$\bullet$} \name improves 10-19\% accuracy and achieves 2-3$\times$ end-to-end throughput compared to the state-of-the-art frame-based enhancement methods. (Sec. \ref{sec:E2EPerf})

\scalebox{0.8}{$\bullet$} \name shows robust effectiveness on various devices with heterogeneous computational resources, different analytical tasks and models, diverse workloads and performance targets, and various resolutions. (Sec. \ref{sec:E2EPerf})

\scalebox{0.8}{$\bullet$} MB-based Region Importance Prediction and Region-aware Enhancement bring remarkable accuracy and throughput benefit, and Profile-based Execution Planning boosts great resource utilization and throughput. (Sec. \ref{sec:ComponentwiseAnalysis})


\vspace{-0.5em}
\subsection{Implementation}
\vspace{-0.5em}
We implement \name upon commercial frameworks including FFmpeg (v4.4.2)~\cite{ffmpeg}, Pytorch (v1.8.2)~\cite{pytorch}, Paddleseg (v2.7.0)~\cite{paddleseg2019}, ONNX, OpenVINO (v2023.0.1)~\cite{openvino}, TensorRT (v8.4.2.4)~\cite{TensorRT}, and the mostly code is implemented by Python (v3.8).
The MB-based region importance prediction and the region-aware enhancer are implemented as follows.
(1) \name implements the MB importance predictor by retraining MobileSeg (MobileNetV2 backbone) with importance metric and prunes its 50\% parameters by L1-Norm pruner. 
It is further exported to the ONNX version (using \texttt{paddle.onnx.export} API) for efficient running on Intel CPU with OpenVINO Runtime, and exported to the TensorRT version for NVIDIA GPU (using \texttt{trtexec} library). 
If not mentioned, all TensorRT models are set as \texttt{FP16} in dynamic shape version, and the engine files are exported and inferred by PyCUDA (v2022.2).
(2) We modify \texttt{ff\_h264\_idct\_add} API in FFmpeg to extract the residual for temporal MB importance reuse.
(3) \name uses a pre-trained super-resolution model \cite{Lim2017EDSR} as enhancer, and also converts it to ONNX version and TensorRT version for efficiency.



Loading models into memory can cost hundreds of milliseconds to seconds. 
Therefore, at runtime, \name pre-loads the DNN into the specified processors and then invokes it.
\name consists of $\sim$5.3K lines of code.

\begin{table}[t] 
\small
\centering 
\begin{tabularx}{0.48\textwidth}{c c|c c}
\toprule 
\textbf{Tasks} & \textbf{Metric} & \textbf{Models} & \textbf{Dataset}  \\
\midrule 
\makecell{Object \\ Detection}  & \makecell{F1-score \\ Stream \#} & \makecell{Mask R-CNN (Swin) \\ YOLO} & \makecell{YODA \\ VideoClips}\\
\midrule 
\makecell{Semantic \\ Segmentation}  & \makecell{mIoU \\ Stream \#} & \makecell{FCN \\ HarDNet} & \makecell{Cityscape \\ BDD100K}\\
\bottomrule 
\end{tabularx}
\caption{Summary of downstream tasks and datasets.} 
\label{tab:downstreamTask}
\vspace{-0.5em}
\end{table}

\vspace{-1em}
\subsection{Experimental Setup}\label{sec:ExperimentalSetup}
\vspace{-0.5em}
\textbf{Downstream tasks and dataset.} 
As summarized in Tab.\ref{tab:downstreamTask}, we select two downstream analytical tasks, object detection and semantic segmentation, to evaluate the performance of \name  
as they play the core role in a wide range of high-level tasks.
Object detection aims to identify objects of interest (\ie, locations and classes) on each video frame; its accuracy is measured by average \emph{F1-score} in each stream
with Intersection over Union (IoU) threshold at 0.5.
Semantic segmentation labels each pixel with one class, and we measure its accuracy with \emph{mIoU}.
We also evaluate their \emph{throughput}, \ie, number of streams that can be processed in real-time.


Both of the tasks were separately tested with two models (light- and heavyweight) on two video sets.
For the object detection, we used the Yoda dataset~\cite{xiao2021towards} and gathered 120 video clips from YouTube containing various scenes with diverse characteristics, \ie, time, illumination, objects' density and speed, and road type. 
The labels to train importance predictor are generated with Mask R-CNN (Swin backbone) on per-frame enhancement since its SOTA performance; if users provide their self-tailored analytical models, we use their output as labels.
Here, we use YOLO as an example.
We will release this dataset after this paper is accepted.
For semantic segmentation, we utilized the BDD100K~\cite{BDD100K} and Cityscape~\cite{Cordts2016Cityscapes} public dataset. 
We re-encoded these datasets into 360P resolution, 30 fps, and 1024kbps bitrate videos in H.264 to avoid the impacts of various video codecs. 
\begin{figure*}[t]
\centering
\setlength{\abovecaptionskip}{0pt}
\begin{minipage}[b]{0.38\linewidth}
    \subfigure[Object detection] {
    \centering     
    \setlength{\abovecaptionskip}{0pt}
    \includegraphics[width=0.46\linewidth]{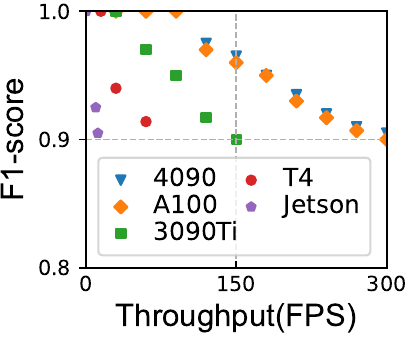}
    }
    \subfigure[Semantic segmentation] {
     \centering
    \includegraphics[width=0.46\linewidth]{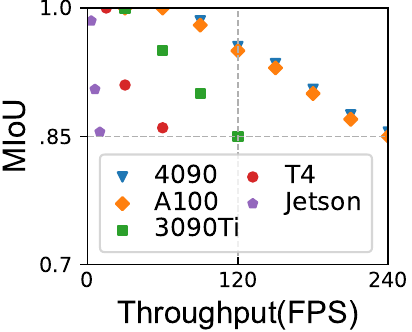}
    }
    \caption{TPT-ACC tradeoff on different devices.}
    \label{fig:Tradeoff}
\end{minipage}
\begin{minipage}[b]{0.61\linewidth}
    \centering
    \setlength{\abovecaptionskip}{0pt}
    \subfigure[Object detection] {
    \centering     
    \includegraphics[width=0.48\linewidth]{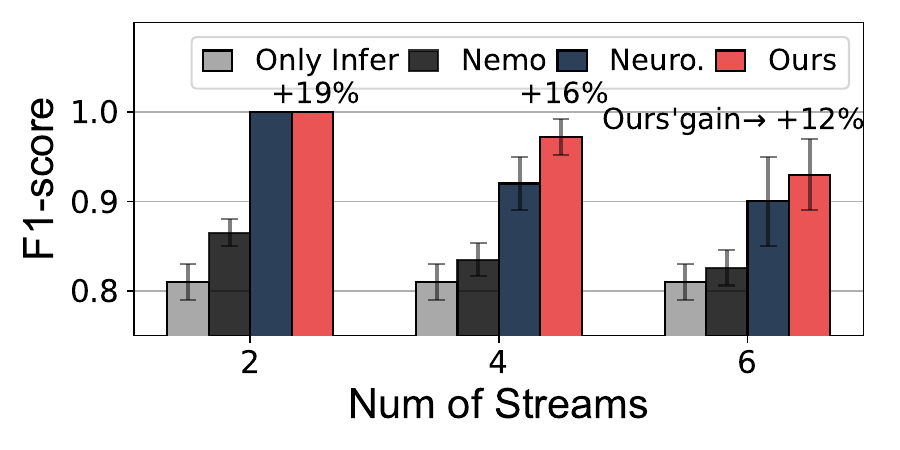}
    }
    \subfigure[Semantic segmentation] {
     \centering
    \includegraphics[width=0.48\linewidth]{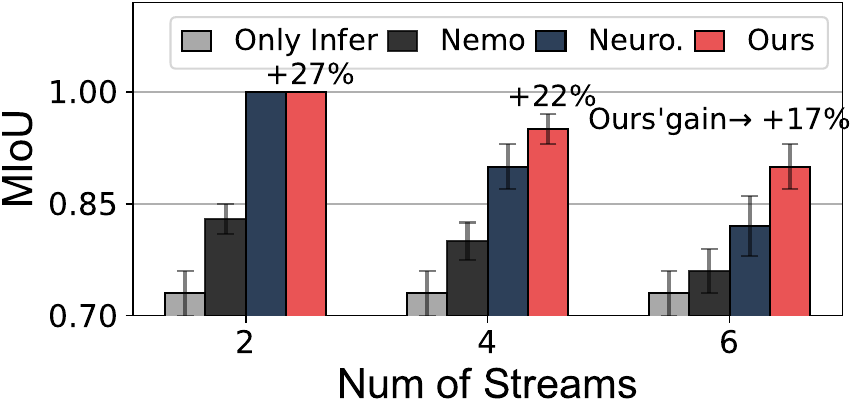}
    }
    \caption{Accuracy comparison over various stream numbers.}
    \label{fig:ACCCompareoverVariousStreams}
\end{minipage}
\vspace{-0.5em}
\end{figure*}

\begin{figure*}[t]
\centering
\begin{minipage}[b]{0.47\linewidth}
    \centering
    \setlength{\abovecaptionskip}{0pt}
    \subfigure{
    \centering     
    \includegraphics[width=0.45\linewidth]{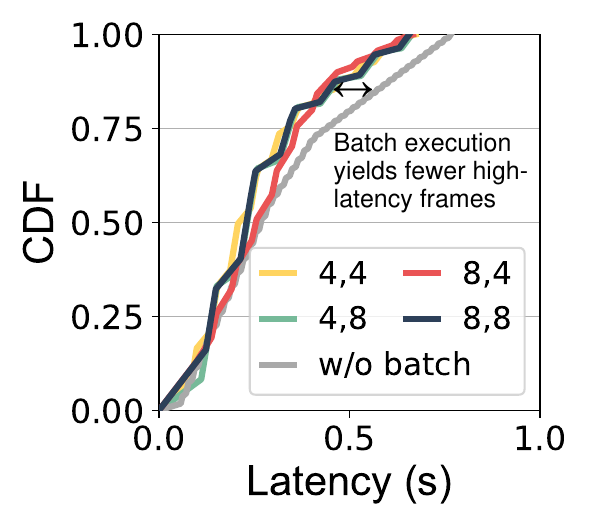}
    }
    \subfigure{
     \centering
    \includegraphics[width=0.45\linewidth]{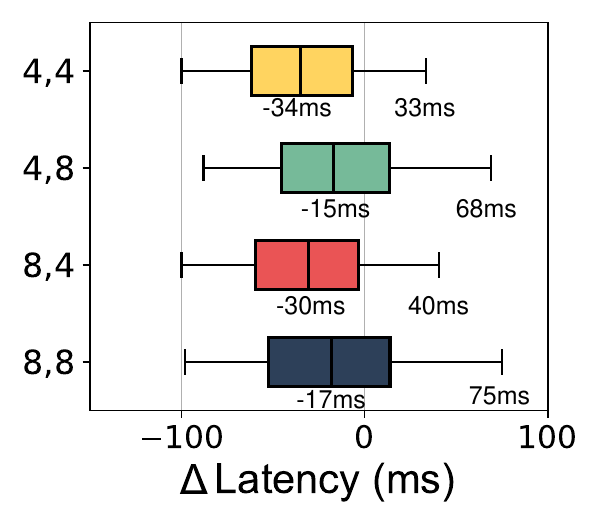}
    }
    \caption{\emph{Left}: Latency of each frame with diverse batch sizes (Infer\#, SR\#). \emph{Right}: The latency diff between execution w/ and w/o batch of each frame ($\Delta Lat=Lat_{batch}(i)-Lat_{w/o\ batch}(i)$).}
    \label{fig:FrameLatency}
\end{minipage}
\hfill
\begin{minipage}[b]{0.24\linewidth}
    \centering
    \footnotesize
    \setlength{\extrarowheight}{2.2pt} 
    \begin{tabularx}{1\textwidth}{c| c c}
      \toprule 
      \textbf{Metric} & \textbf{360P} & \textbf{720P}\\
      \midrule 
        BW (Mbps)  & 0.96 & 3.00\\
      \midrule 
       Max Stream \# & 11 & 10\\
      \midrule 
      GPU Usage (SR)  & 0.23 & 0.17\\
     \midrule 
      Acc Gain & 9\% & 7\%\\
     \bottomrule 
    \end{tabularx}
    \captionof{table}{Performance trade-off under different video resolutions.}
    \label{fig:BW}
  %
\end{minipage}
\hfill
\begin{minipage}[b]{0.24\linewidth}
    \centering
    \footnotesize
    \begin{tabularx}{1\textwidth}{c c}
      \toprule 
      \textbf{Method} & \textbf{TPT (FPS)}\\
      \midrule 
        Per-frame SR  & 95\\
      \midrule 
      PF + Plan.  & 111\\
      \midrule 
      PF + Pred. + Plan.  & 111\\
     \midrule 
      PF + Pred. + Enhan. & 179\\
      \midrule 
        \name & 300\\
     \bottomrule 
    \end{tabularx}
    \captionof{table}{End-to-end throughput (FPS) breakdown of \name.}
    \label{fig:com_breakdown}
\end{minipage}
\vspace{-1em}
\end{figure*}

\noindent
\textbf{Devices.} 
We conducted experiments on five heterogeneous devices divided into four categories.
(1) As a comparison, we deployed and tested RegenHance on a cloud server with an NVIDIA A100 GPU and Intel(R) i9-12900K CPU.
(2) As one of the most popular configurations of edge servers, we conducted experiments on an edge equipped with an NVIDIA Tesla T4 and Intel i7-8700 CPU.
(3) To explore the capabilities of gaming cards and their generation gap, we tested an NVIDIA RTX4090 and an NVIDIA RTX3090Ti, respectively, with an Intel i9-13900K.
(3) For the embedded edge, we used an NVIDIA Jetson AGX Orin 64GB as the platform.

\noindent
\textbf{Baselines.} To show the superiority brought by \name, we compare it with four different baselines:\\
\scalebox{0.8}{$\bullet$} 
\textbf{Only infer} directly applies analytical DNNs on each original frame without enhancement. \\ 
\scalebox{0.8}{$\bullet$} 
\textbf{NeuroScaler~\cite{yeo2022neuroscaler}} is the state-of-the-art frame-based enhancement method. 
It first enhances the anchors and reuses their quality gain on non-anchors, then infers all frames. 
It fast selects anchors in a heuristic manner.\\
\scalebox{0.8}{$\bullet$} 
\textbf{Nemo~\cite{yeo2020nemo}} also only enhances the anchors and reuses their quality gain, then infers all frames; but it iteratively selects the best anchors based on enhancement results.

\vspace{-0.5em}
\subsection{End-to-End Performance}\label{sec:E2EPerf}
This section reports the E2E performance of \name on various devices.
If not mentioned, all results are tested on RTX4090 with 1s latency, 90\% object detection, and 88\% semantic segmentation (10\% gain compared to only infer).


\noindent
\textbf{Performance on heterogeneous devices.} 
As shown in Fig. \ref{fig:odDevicecompare} and Fig. \ref{fig:ssDevicecompare}, \name achieves high accuracy and high throughput simultaneously on various devices.
It of course cannot reach the throughput of only infer due to the additional enhancement time cost but provides significant throughput gain compared to SOTA NEMO and NeuroScale. 
On average, \name outperforms their throughput by 12$\times$ and 2.1$\times$ in object detection and 11$\times$ and 1.9$\times$ in semantic segmentation, respectively. 
This larger performance gain in semantic segmentation stems from its heightened sensitivity to visual details.
\name can keep this superiority on all five heterogeneous devices because profile-based execution planning always generates the optimal throughput plans.

\noindent
\textbf{Trade-off between throughput and accuracy.} 
\name creates a trade-off space between accuracy and throughput, as shown in Fig. \ref{fig:Tradeoff}; it can offer service on edge servers with heterogeneous resources.
Edge servers equipped with higher resources produce larger trade-off space.
For example, \name supports object detection for ten streams (300fps) with 91\% accuracy on RTX4090 
or A100 GPU; 
and if the accuracy constraint is more strict, \name will make corresponding adjustments and serve the maximum number of streams, \eg, six 95\%-accuracy streams.
On devices with lower available resources, \eg, on NVIDIA T4 and Jetson AGX Orin, although the maximum frame rate decreases, \name still delivers remarkable throughput under different accuracy targets.




\noindent
\textbf{Performance gain with multiple streams.} 
With the increasing number of competing streams, \name always achieves the highest accuracy compared to the other three frame-based enhancement methods.
For example, in Fig. \ref{fig:ACCCompareoverVariousStreams} tests on RTX4090, \name improves accuracy by 8-14\% compared to the selective enhancement in six streams. 
This is because, in such a highly competitive scenario where each stream is allocated limited resources, our method consistently enhances the most valuable regions; in contrast, the selective baseline and the per-frame baseline waste excessive resources on unimportant content.
\name significantly improves the throughput of content-enhanced video analytics.

\begin{figure*}[t]
\begin{minipage}[t]{0.19\linewidth}
\centering
\includegraphics[width=1.\linewidth]{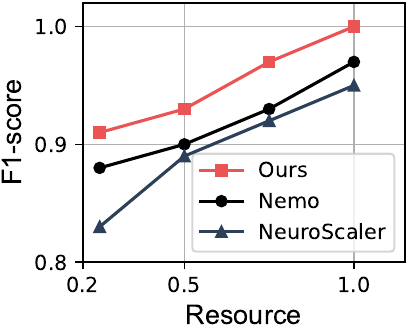}
\vspace{-2em}
\caption{Accuracy on different resources.}
\label{fig:s_bub-a}
\end{minipage}
\hfill
\begin{minipage}[t]{0.185\linewidth}
\centering
\includegraphics[width=1.\linewidth]{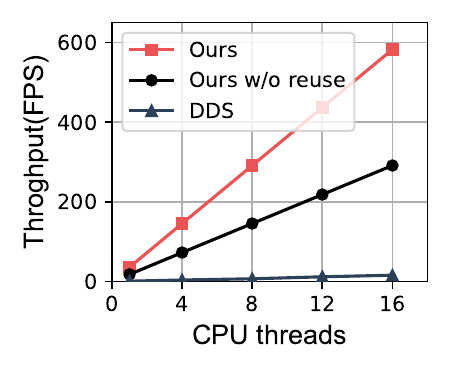}
\vspace{-2em}
\caption{Throughput of region prediction.}
 \label{fig:s_bub-b}  
\end{minipage}
\hfill
\begin{minipage}[t]{0.185\linewidth}
\centering
\includegraphics[width=1.05\linewidth]{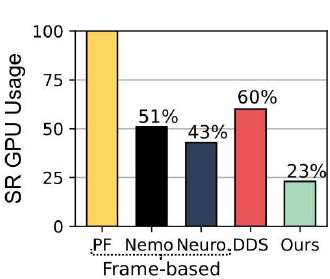}
\vspace{-2em}
\caption{GPU resources usage.}
\label{fig:s_bub-c}  
\end{minipage}
\hfill
\begin{minipage}[t]{0.2\linewidth}
\centering
\includegraphics[width=1.03\linewidth]{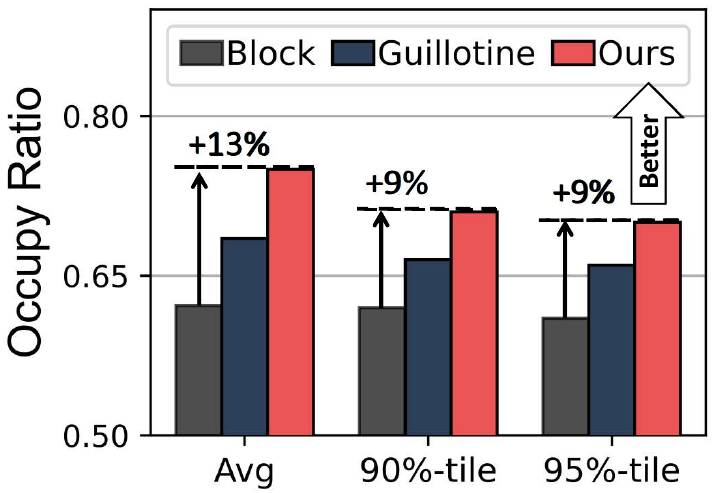}
\vspace{-2em}
\caption{Perf. of different packing policies.}
\label{fig:s_bub-d}  
\end{minipage}
\begin{minipage}[t]{0.19\linewidth}
    \centering
    \includegraphics[width=1.1\textwidth]{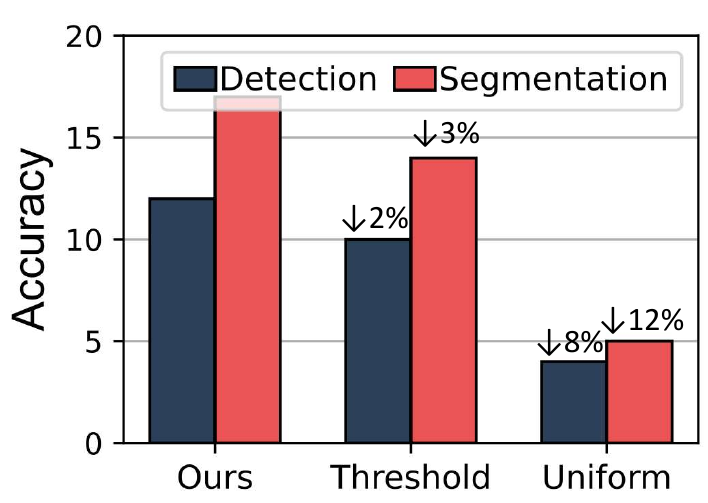}
    \vspace{-2em}
    \caption{Acc gain of Cross-stream MB sel.}
    \label{fig:breakdown_sch}
\end{minipage}
\vspace{-1em}
\end{figure*}

\noindent
\textbf{Frame latency under diverse batch sizes.} 
The latency is defined the same as prior studies (\eg, DDS \cite{du2020server}, AWStream~\cite{zhang2018awstream}, and Reducto \cite{li2020reducto}), \ie, the time from encoding 1s video chunk (30 frames) on cameras to all 30 frames' inference results are available on edge. 
The left figure of Fig. \ref{fig:FrameLatency}  plots the latency of each frame with different batch sizes under the same workflow.
The results imply that batch execution yields fewer high-latency frames because its average latency of 30 frames in each chunk is much lower than executed without batch, which stems from the batch's higher GPU utilization.
For a deeper insight, the right figure of Fig. \ref{fig:FrameLatency} evaluates the latency difference of each frame with and without batch execution.
Compared to without batch, execution with batch may cause 75ms latency at most, which is the earliest frame (in video play order) in one batch; but on average, batch execution saves time cost.

\noindent
\textbf{Performance gain under different resolutions.}
When streaming 720p in object detection while 360p for semantic segmentation under a target of 90\% accuracy, as shown in Table \ref{fig:BW}, 360p video costs only 31\% bandwidth (0.96Mbps \vs 3Mbps) compared to 720p.
\name boosts accuracy gain of 360p video from 81\% to 90\%, while 720p video from 83\% to 90\%.
Higher resolution yields higher accuracy; however, content enhancement can still improve the details of small objects or blurred content.
The end-to-end throughput of 720p is almost the same as 360p (11 \vs 10 streams). 
The reason is that although 720p needs to enhance fewer regions compared to 360p (17\% \vs 23\% GPU usage for SR), its time cost of other runtime components (\eg, MB importance prediction) is larger than 360p due to the larger input size.

\noindent
\textbf{Contributions by individual components.} 
Table \ref{fig:com_breakdown} shows the throughput breakdown of \name. 
We tested multiple versions of \name by selectively applying each component.
The results show that each component significantly contributes to the improvements in the throughput.
The execution planning enables a 1.2$\times$ throughput compared to per-frame enhancement by reasonable resource allocation (1st→2nd).
Incorporating MB-based region importance prediction without region-aware enhancement fails to improve throughput because filling unimportant regions to black does not alter the enhancement latency (2nd→3rd row).
The throughput is increased by 1.6$\times$ due to the region-aware enhancement (3rd→4th row) and by 1.7$\times$ from the best execution plan of profile-based execution planning (4th→5th row).

\subsection{Comprehensive Component-wise Analysis}\label{sec:ComponentwiseAnalysis}
We provide an in-depth performance analysis of individual system components.

\subsubsection{MB-based region Importance Prediction}\label{sec:ExperimentPrediction}(\S \ref{sec:RegionSelector}) helps \name achieve great throughput improvement while keeping high accuracy.


\noindent
\textit{Accuracy:} 
Fig. \ref{fig:s_bub-a} illustrates the accuracy gain obtained by different enhancement methods, when assigned the same computational resource to complete object detection on six streams. 
\name (region-based enhancement) can offer a 3-4\% and 4-8\% higher gain compared to (frame-based) Nemo and Neuralscaler, respectively, because our predictor identifies the most beneficial region precisely. 


\noindent
\textit{Thoughput:} 
As shown in Fig. \ref{fig:s_bub-b}, our ultra-lightweight MB importance predictor can be executed at 30 fps on one single i7-8700 CPU core, which outperforms the RoI selection by RPN in DDS more than 60 times.
On GPU, it achieves 973 fps, which outperforms DDS more than 12 times.
The reuse contributes 2 times throughput improvement.


\noindent
\textit{Resource saving:} 
Fig. \ref{fig:s_bub-c} shows that our method reduces $77\%$, $28\%$, and $20\%$ GPU usage compared to the frame-based Per-frame enhancement, Nemo, and NeuroScaler, respectively, when enhancing a single stream (30FPS) to achieve an accuracy exceeding 90\%. 
Compared to the RoI selection of DDS, our method saves 37\% GPU usage as it identifies the most beneficial region for analytical accuracy more precisely.

\subsubsection{Region-aware Enhancement}\label{sec:ExperimentEnhancement}(\S \ref{sec:EnhancementExecutor}) yields significant throughput improvement and overall accuracy gain. 

\noindent
\textit{Throughput:} 
Region-aware bin packing maximizes \name's throughput
with a high stability and occupy ratio, and thus effectively reducing the additional overhead of SR.
We conducted experiments repeated 1,000 times by randomly shuffling the order of six video streams, and comparing the occupy ratio, \ie, the ratio of selected MBs occupying all enhanced content, with the classic Guillotine policy \cite{puzzle} and the Block policy, \ie, MB packing. 
Fig. \ref{fig:s_bub-d} shows that the average, 90\%-tile, and 95\%-tile occupy ratio difference; our packing policy gets the highest occupy ratio of 75\% that 
outperforms the comparisons by up to 13\%, 9\%, and 9\%. 

\begin{figure*}[t]
\centering
\begin{minipage}[b]{0.18\linewidth}
    \centering
    \includegraphics[width=0.95\textwidth]{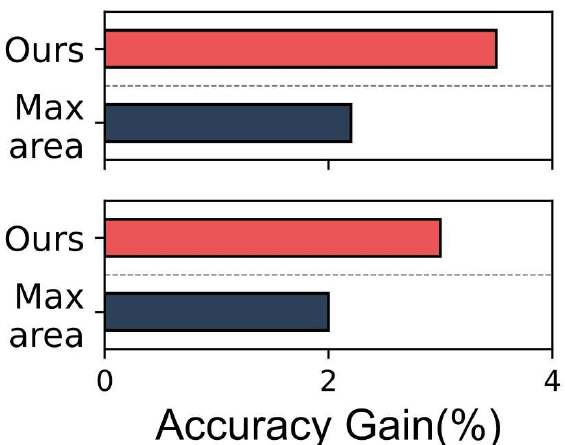}
    \caption{Acc gain of our MB selection.}
    \label{fig:area_score}
\end{minipage}
\hfill
\begin{minipage}[b]{0.20\linewidth}
    \centering
    \includegraphics[width=0.95\textwidth]{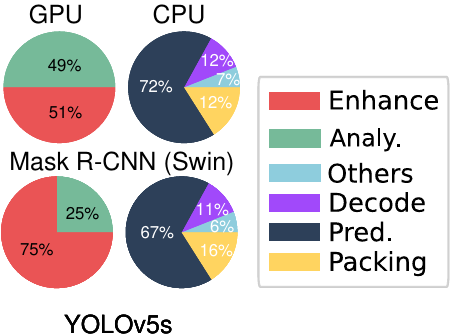}
    \caption{Execution plan for different workload.}
    \label{fig:PlanWorkload}
\end{minipage}
\hfill
\begin{minipage}[b]{0.34\linewidth}
    \centering
\includegraphics[width=0.95\linewidth]{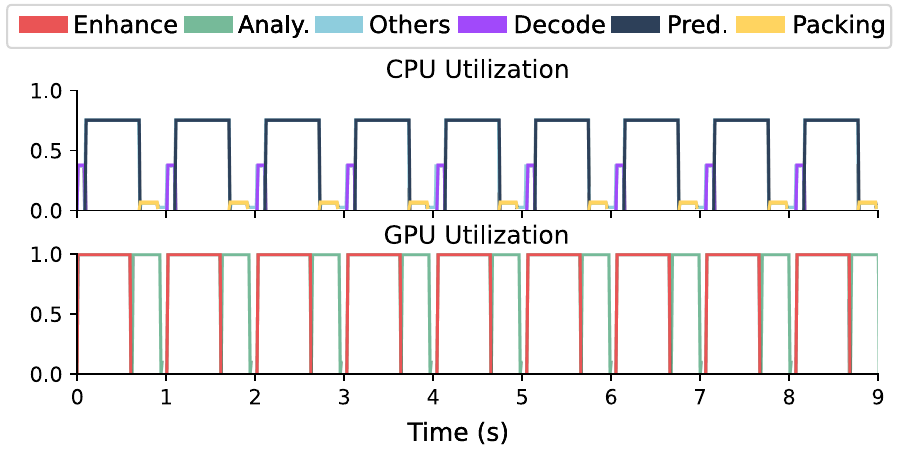}
\caption{GPU \& CPU usage.}
\label{fig:c_bub-a}
\end{minipage}
\hfill
\begin{minipage}[b]{0.26\linewidth}
    \footnotesize
    \centering
    \begin{tabularx}{0.95\linewidth}{c|c|c}
    \toprule 
    \textbf{Component} & \textbf{RRobin} & \textbf{Ours}  \\
    \midrule 
    \makecell{MB Prediction}  & \makecell{194} & \makecell{533}\\
    \midrule 
    \makecell{Enhancement}  & \makecell{80}  & \makecell{186}\\
    \midrule 
    \makecell{Analytics}  & \makecell{220} & \makecell{186}\\
    \midrule 
    \makecell{\textbf{Throughput}}  & \makecell{\textbf{80}}  & \makecell{\textbf{186}}\\
    \bottomrule
    \end{tabularx}
    \captionof{table}{Throughput (FPS) gain to region-agnostic strawman in \S \ref{sec:challenges}.}
    \label{fig:c_bub-b}  
\end{minipage}
\vspace{-1em}
\end{figure*}

\noindent
\textit{Accuracy:} 
Cross-stream MB selection, especially our custom sorting order, leads to remarkable accuracy gain by considering the heterogeneous region accuracy gain across streams. 
We compared our MB selection with the Uniform method, which allocates the same MB number to each stream, and the Threshold method sets a fixed threshold, 0.5, to select MB importance for all streams.
As shown in Fig. \ref{fig:breakdown_sch}, our method outperforms the Uniform and Threshold methods by 8-12\% and 2-3\% accuracy gain, respectively.
Fig \ref{fig:area_score} demonstrates the superiority of our custom sorting priority.
Compared to the traditional large-item-first (max-region-first in our context) packing policy, our method yields a 50\% accuracy gain.

\subsubsection{Profile-based Execution Planning }\label{sec:ExperimentPlanning}(\S \ref{sec:ComponentResourceManager}) profiles different devices and allocates devices and resources to different components and models based on the results, thereby maximizing throughput.

\noindent
\textit{Resources allocation:} 
Profile-based execution planning makes \name achieve the best performance on given devices and workloads by avoiding bottlenecked by any component. 
Fig. \ref{fig:PlanWorkload} visualizes the computational resource allocation of two different object detection models on i9-13900K+RTX4090 providing satisfied performance as Fig. \ref{fig:4090}.
With distinct workloads of YOLOv5s (16.9GFLOPs) and Mask R-CNN (Swin backbone, 267GFLOPS), it allocates more resources for the analytical task.

\noindent
\textit{Resource usage:} Fig. \ref{fig:c_bub-a} shows the real-time utilization of CPU and GPU when detecting objects on six video streams.
We used Nvidia Nsight to monitor GPU utilization and HTOP for CPU utilization.
The GPU can reach full load approximately 95-99\% of the time, while the CPU can reach high load around 81\% of the time. 
This result indicates that the execution plan achieves an efficient GPU-CPU corporation. 

\noindent
\textit{Throughput:} 
We avoid the enhancement bottleneck introduced by the round-robin manner (\S \ref{sec:challenges}), which allocates equal resources to each component.
The results in Table \ref{fig:c_bub-b} indicate that our approach achieves 2.3$\times$ throughput. 

\noindent
\textit{Scalability and initialization:}
There are two types of preparation time in our system. 
If only the ingest video streams change, it needs approximately 0.6-2 seconds for initialization depending on the specific device and model requirements.
When given a new device, it needs to take approximately 1-3 minutes to execute the Profile-based Execution Planning.

\vspace{-0.5em}
\section{Related Work}\label{sec:RelatedWorks}
\vspace{-0.5em}

\textbf{Video analytics} is facilitated by the advances of deep learning.   
However, the high analytical accuracy from deep learning comes at a prohibitive computing cost that most end devices cannot afford~\cite{Wu2019Facebook}. 
For example, today's deployed traffic cameras cost only \$40 to \$200, equipped with a single-core CPU that provides very scarce computational resources.
To tackle this issue, today's video analytics apply a \emph{distributed} architecture~\cite{zhang2015design, kang2017noscope, meddebROI, zhang2021elf, hsieh2018focus, liu2019edge, du2022AccMPEG, li2020reducto, yuan2022infi, AccDecoder, yao2020deep, yuan2022PacketGame}.

\textbf{Video compression} saves bandwidth resources by adjusting video encoding configurations (\eg, resolution, bitrate, and frame rate). 
Like traditional video streaming protocols like WebRTC~\cite{WebRTC} and DASH~\cite{stockhammer2011dynamic}, distributed video analytics must tackle bandwidth constraints.
Prior studies, no matter server-feedback accurate configuration setup~\cite{zhang2018awstream, jiang2018chameleon, meddebROI, zhang2021elf} or fast autonomy by video source (camera) itself~\cite{liu2019edge, yao2020deep, du2022AccMPEG, xie2019source}, addressed this issue well, saving considerable bandwidth resources but causing accuracy drop.

\textbf{Selective inference}, also called frame/image filtering, filters similar frames/images out to meet low resource cost and high throughput.
Cameras leverage pixel-level difference~\cite{chen2015glimpse, li2020reducto, dai2021cinet} to filter frames for bandwidth saving; edge servers use light DNNs to select the frames to be detected by heavy DNNs, then reuse their output on filtered frames for throughput increment~\cite{yuan2022infi, AccDecoder, kang2017noscope, zhang2015design, hsieh2018focus}.
Selective inference achieves satisfactory throughput and resource-saving but lower accuracy. 
DDS~\cite{du2020server} iteratively detects the selected images to avoid accuracy drop, but its twice cross-network transmission and detection imports too much delay.
Turbo~\cite{lu2022turbo} seeks the same goal by enhancing frames on idle GPU slots. 
Differing from it, \name boosts the enhancement method and fully uses heterogeneous resources on edge.
\textbf{Model optimization} meets high throughput by network pruning~\cite{li2016pruning} and weights quantization~\cite{jacob2018quantization}, and achieves high accuracy by knowledge distillation~\cite{hinton2015distilling}.
However, the compressed model is easily affected by data drift, where the live video data diverges from the training data, 
leading to accuracy degradation~\cite{Romil2022Ekya}.
To tackle this, continuous learning~\cite{Romil2022Ekya, kim2020neural}, model switching~\cite{khani2023recl, yeo2022neuroscaler}, and model merging~\cite{padmanabhan2022gemel, Jiang2018Mainstream} are proposed by previous studies.
Intuitively, continuous learning and model switching leverage small DNNs for high throughput and improve accuracy by matching the weights in DNN to the changed real-world videos. 
Model merging differs from traditional parameter sharing in one model~\cite{pham2018efficient}. 
It explores structural similarities between concurrent models and enables multiple models to share the same network components, \eg, the backbone~\cite{padmanabhan2022gemel}, by retraining. 


\vspace{-1em}
\section{Conclusion}
\vspace{-0.5em}
In this paper, we look at content enhancement in video analytics applications.
We found that frame-based content enhancement wastes too much computation on the analytics-agnostic image. 
We presented the region-based content enhancement technique and well-matched region-aware resource scheduler and implemented \name.
In our evaluation using five heterogeneous devices, we show that \name can deliver an order of magnitude than frame-based content-enhanced video analytics.

\vspace{0.5em}
\noindent
\textbf{Acknowledgements.}
We thank the anonymous NSDI reviewers for their constructive comments. 
This work was supported by Carbon Neutrality and Energy System Transformation (CNEST) Program,  Tsinghua University (AIR)-AsiaInfo Technologies (China), Inc. Joint Research Center for 6G Network and Intelligent Computing, NSFC (No. 62272261, 
62402280, 
U22A2031) 
the Fundamental Research Funds for the Central Universities (No. 2024300349), 
EU Horizon CODECO projects (No. 101092696),
Shuimu Tsinghua Scholar Program (No. 2023SM201).


\newpage
\bibliographystyle{plain}
\bibliography{reference.bib}

\newpage
\appendix

\section{Functions invoked in Algorithm \ref{alg:binpacking}}\label{sec:FunctionsinAlg}
\begin{algorithm}\footnotesize
\caption{InnerFree(farea,box)}
\label{alg:InnerFree}
\begin{algorithmic}[1]
\algnotext{EndIf}
\algnotext{EndFor}
\algnotext{EndFunction}
\Require farea,box
\Ensure the rest free area in box
\Function{InnerFree}{farea, box}  
\State m,n,dp=box.w,box.h,List[m][n]
\State \Call{Init}{dp} 

\For {j \textbf{in} n}
    \State up,down,stk=List[m],List[m],Stack()
    \For {i \textbf{in} m}
        \If{stk is not None and left[stk.top()][j] >= left[i][j]} 
        \State down[stk.pop()] = i
        \EndIf
        \If{stk is not None}
        \State up[i] = stk.top()
        \EndIf
        \State stk.push(i)
    \EndFor
    \For {i \textbf{in} m}
        \State height = down[i] - up[i] - 1
        \State area = height * left[i][j]
        \State get the free area with max area
    \EndFor
\EndFor
\State \textbf{return} max free area
\EndFunction

\end{algorithmic}
\end{algorithm}

\section{Performance of Importance Level Approximation}\label{sec:Derivation}
\noindent
\begin{figure}[h]
  \centering
    \includegraphics[width=0.55\linewidth]{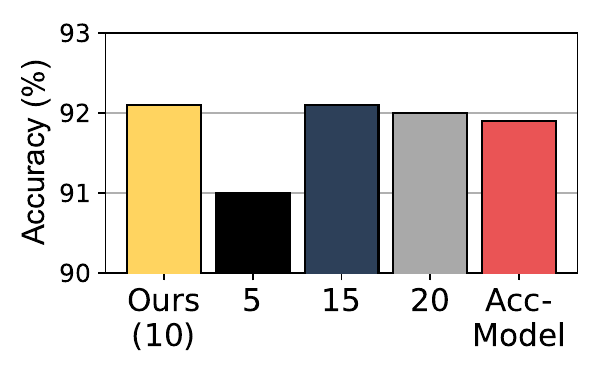}
    \caption{Importance level classification of MBs (\ie, MB-grained image segmentation assigns each MB an importance level) yields comparable (even better) accuracy than AccModel, which predicts exact importance value.}
    \label{fig:lossfunc2}
\end{figure}

\noindent
Approximating the MB importance prediction as an image segmentation problem to assign each MB an importance level (class) is effective and produces good performance.
We have trained four MB importance predictors (in MobileSeg architecture) with 5, 10, 15, and 20 important levels, respectively.
Fig. \ref{fig:lossfunc2} compares the final accuracy of inference task when predicting the exact MB importance (AccModel) and important levels.
Experiment results demonstrate that important level classification yields precise MB importance prediction and hence high accuracy provided the level number is not very coarse, saying 5.
The efficiency of this importance level classification is because such shallow model architecture is better at more manageable classification tasks (\eg, segmentation) than regression tasks (\eg, AccModel predicts the exact importance value).
\name set level number as 10.

\section{Additional Results}\label{sec:AddtionalResults}
\subsection{Example Eregions and the Distribution}\label{sec:AdditionalResults1}
\begin{figure}[h]
     \centering
    \includegraphics[width=0.4\textwidth]{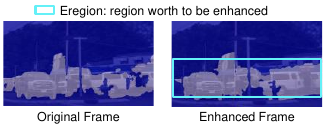}
    \caption{Example eregion worth to be enhanced for semantic segmentation (SS) bounding with a rectangle box.}
    \label{fig:ss_eregion}
    \vspace{-0.5em}
\end{figure}

Example eregions of object detection in \S \ref{sec:OurChoice} and semantic segmentation in Fig. \ref{fig:ss_eregion} are generated by simply bounding the different analytical results between original and enhanced frames with a rectangle.
Note that these example eregions are not the eregions used in \name as they still contain much unnecessary content; this is why we propose to set fine-grained MB as the construction unit of eregions in \name.

\begin{figure}[h]
     \centering
    \includegraphics[width=0.27\textwidth]{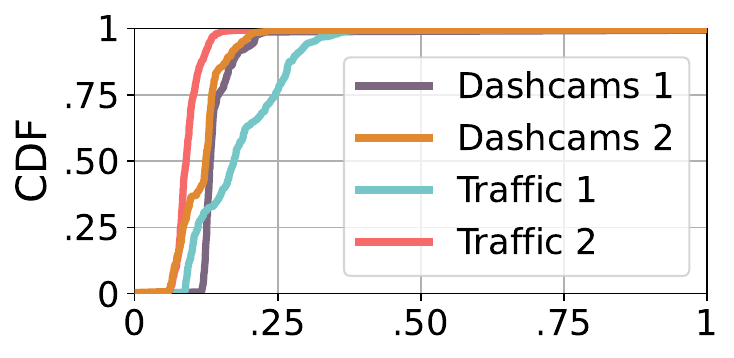}
    \caption{Distribution of eregions in Semantic segmentation.}
    \label{fig:ss_cdf}
    \vspace{-0.5em}
\end{figure}

In semantic segmentation, as illustrated in Fig. \ref{fig:ss_cdf}, only 10-15\% area in 70\% frames are eregions.

\subsection{Performance of Diverse Operators}\label{sec:AdditionalResults2}
 \vspace{-1em}
\begin{figure}[h]
\centering
    \subfigure[CNN \cite{zhi2021mgsampler}] {
    \centering     
    \includegraphics[width=0.2\textwidth]{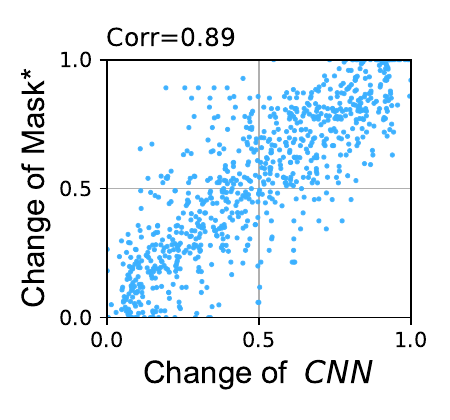}
    }
    \hfill
    \subfigure[Edge] {
     \centering
    \includegraphics[width=0.2\textwidth]{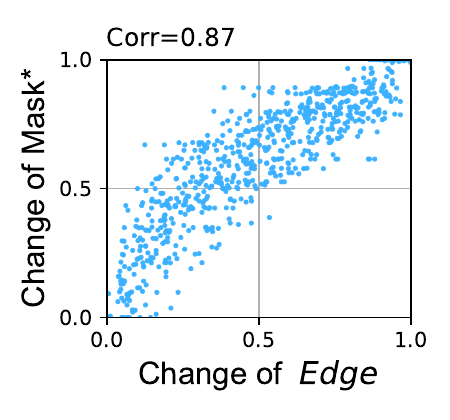}
    }
    \caption{Correlation ratio between the change of features.}
    \label{fig:correlationappend}
    \vspace{-1em}
\end{figure}

\begin{figure}[h]
\centering
    \subfigure[$\frac{1}{Area}$: 0.002, Area: 0.665] {
    \centering     
    \includegraphics[width=0.45\textwidth]{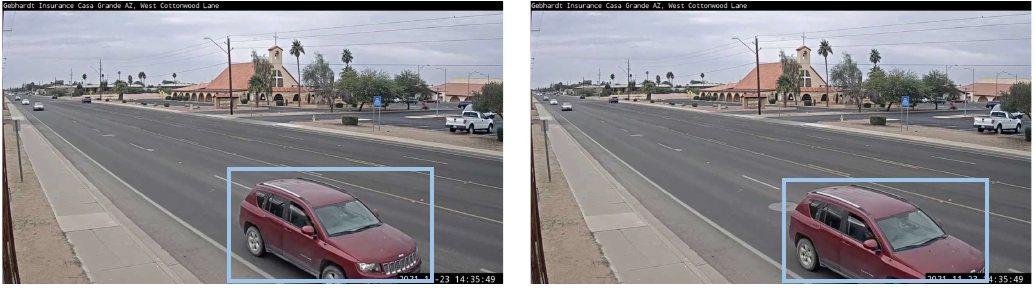}
    }
    \hfill
    \subfigure[$\frac{1}{Area}$: 0.307, Area: 0.035] {
     \centering
    \includegraphics[width=0.45\textwidth]{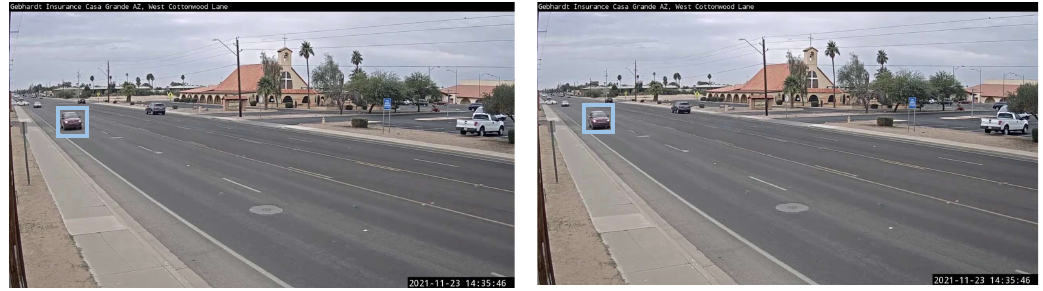}
    }
    \caption{Correlation ratio between the change of features.}
    \label{fig:area&area_1}
\end{figure}

\noindent
As shown in Fig. \ref {fig:correlationappend}, one-layer CNN~\cite{zhi2021mgsampler} and the Edge operator (\ie, edge detector in computer vision community) offer lower correlation ratios compared to the $\frac{1}{Area}$ operator, \ie, 0.91.
Fig. \ref{fig:area&area_1} exemplifies the characteristics of Area and $\frac{1}{Area}$.
Images in the top raw demonstrate that the Area operator captures the change of large blocks, while the values of $\frac{1}{Area}$ change very small.
On the contrary, $\frac{1}{Area}$ captures the change of small objects in two bottom images as \name needs.

\subsection{Performance \vs Expanding Pixels}\label{sec:AdditionalResults3}
\begin{figure}[h]
     \centering
    \includegraphics[width=0.31\textwidth]{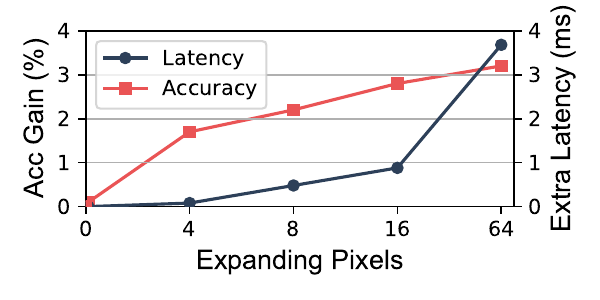}
    \caption{Both accuracy gain and enhancement cost increase with more expanding pixels. \name expands 3 pixels in each direction surrounding every region.}
    \label{fig:pixel_expand}
    \vspace{-0.5em}
\end{figure}

Pixel expanding in each direction can avoid the MB/region boundaries causing too many jagged edges and blocky artifacts when pasting enhanced content back to the bi-linear-interpolated frames. 
But it also introduces extra enhancement costs.
We measure the accuracy gain and extra latency of the object detection task, as shown in Fig. \ref{fig:pixel_expand}.
To balance the accuracy and the enhancement cost, we expand three pixels in this paper because the MB selection (\S 3.3.1) can control the number of enhanced MB to adjust the performance of \name. 

\subsection{Performance of MB Packing and Irregular Packing}\label{sec:AdditionalResults4}
\begin{figure}[h]
     \centering
    \includegraphics[width=0.29\textwidth]{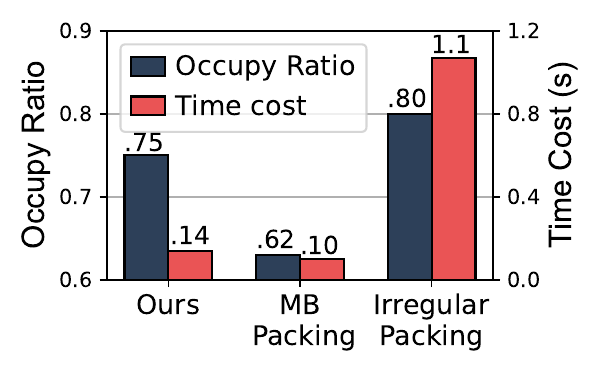}
    \caption{Our packing algorithm achieves a very good balance between bin utilization (occupy ratio) and the time cost of packing plan search.}
    \label{fig:MBIrregularPacking}
    \vspace{-0.5em}
\end{figure}
Our region-aware packing algorithm spends almost the same time as MB packing and provides a close occupy ratio of irregular packing.
As shown in Fig. \ref{fig:MBIrregularPacking}, MB packing yields a low occupy ratio as packing too many unimportant or repeated pixels, while irregular packing leads to a large time cost of packing plan search.



\subsection{An Example of Intermediate Results of RegenHance in Object Detection Task}\label{sec:AdditionalResults5}
\begin{figure*}[t]
     \centering
    \includegraphics[width=0.75\linewidth]{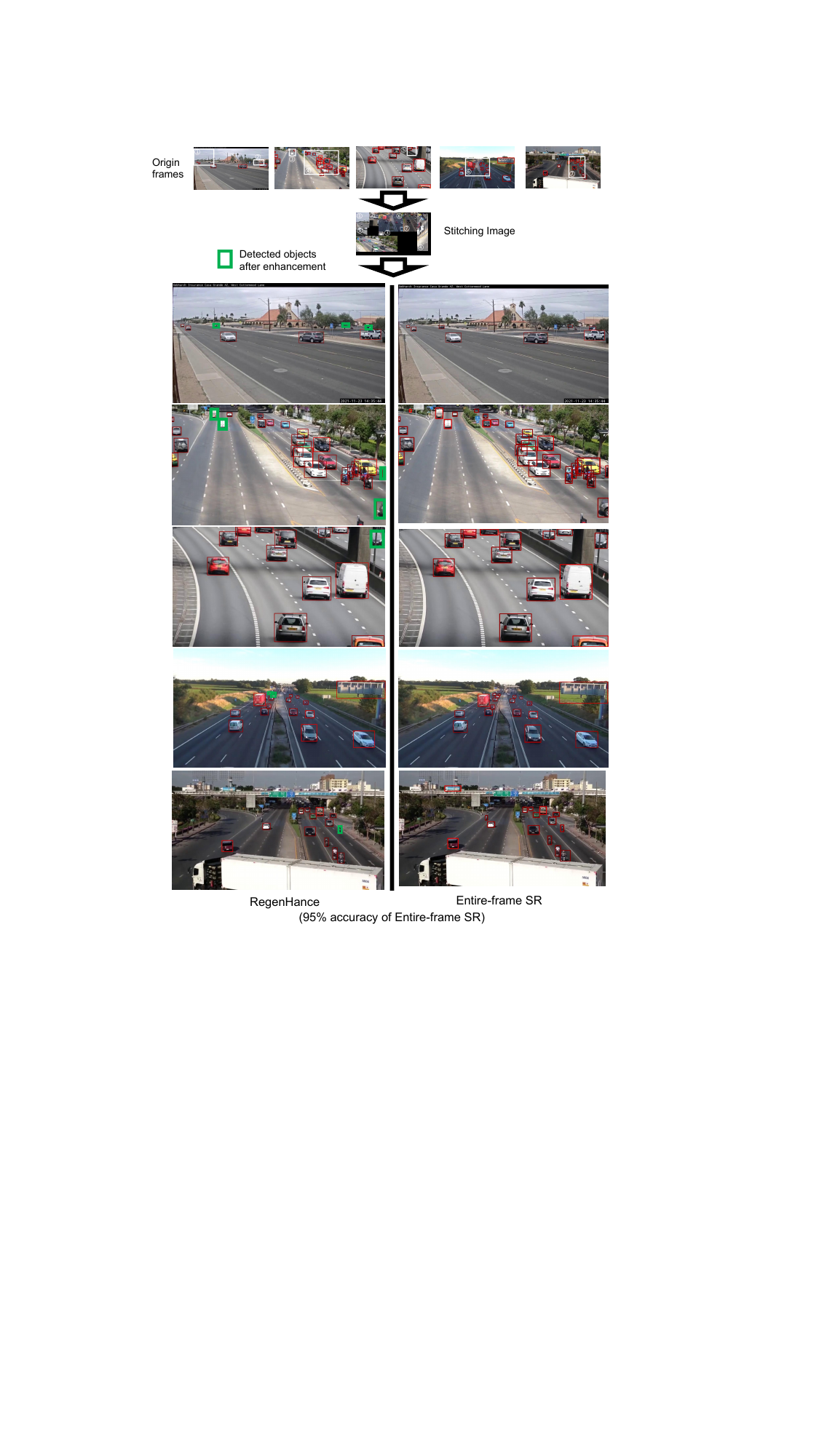}
    \caption{Examples of running super-resolution on stitched regions and their inference results compared to entire-frame enhancement.}
    \label{fig:stitchinig}
\end{figure*}

Fig. \ref{fig:stitchinig} shows an example of packing five frames into one bin and pasting enhanced regions in the stitching image back to the interpolated frames.
In particular, the region \ding{175} is cut into two smaller ones by our region-aware bin-packing algorithm for better bin utilization.
The green boxes in the left column bound the objects can be detected after enhancing the stitching image, \ie, enhancing the selected regions.
To the right column, region-based enhancement offers comparable accuracy with entire-frame enhancement. 

\subsection{Performance under Various Workloads and Users' Latency Targets}\label{sec:AdditionalResults6}
\begin{figure}[h]
    \centering
    \includegraphics[width=0.8\linewidth]{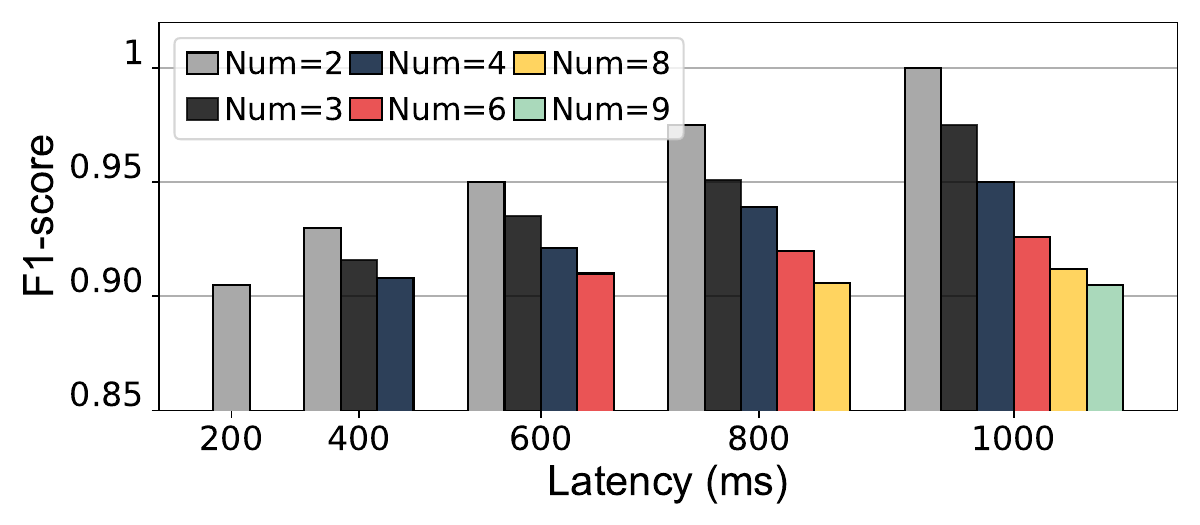}
    \caption{\name supports various user-specified latency targets with adaptive batch sizes under different workloads (\ie, stream numbers)}
    \label{fig:delay}
\end{figure}

\noindent
\name supports various latency targets by automatically adjusting every component's batch size.
Fig. \ref{fig:delay} demonstrates that
\name can well analyze two 30-fps streams (60 frames) under 200ms latency budget and nine streams under 1s.
The batch sizes of components change under different workloads. 
For example, under a 400ms budget, when stream numbers increase from 2 to 4, the batch sizes of (enhanced model, analytical model) are (8,4), (4,8), and (4,8); this implies more resources are allocated from region enhancement to final inference when workload increases, and hence lower accuracy.
Note that the batch sizes of any components are no more than 8 among all latency targets; namely, in each batch, the earliest input won’t wait for the last one more than 75ms as shown in Fig. \ref{fig:FrameLatency}.

\begin{figure*}[t]
     \centering
    \includegraphics[width=1\linewidth]{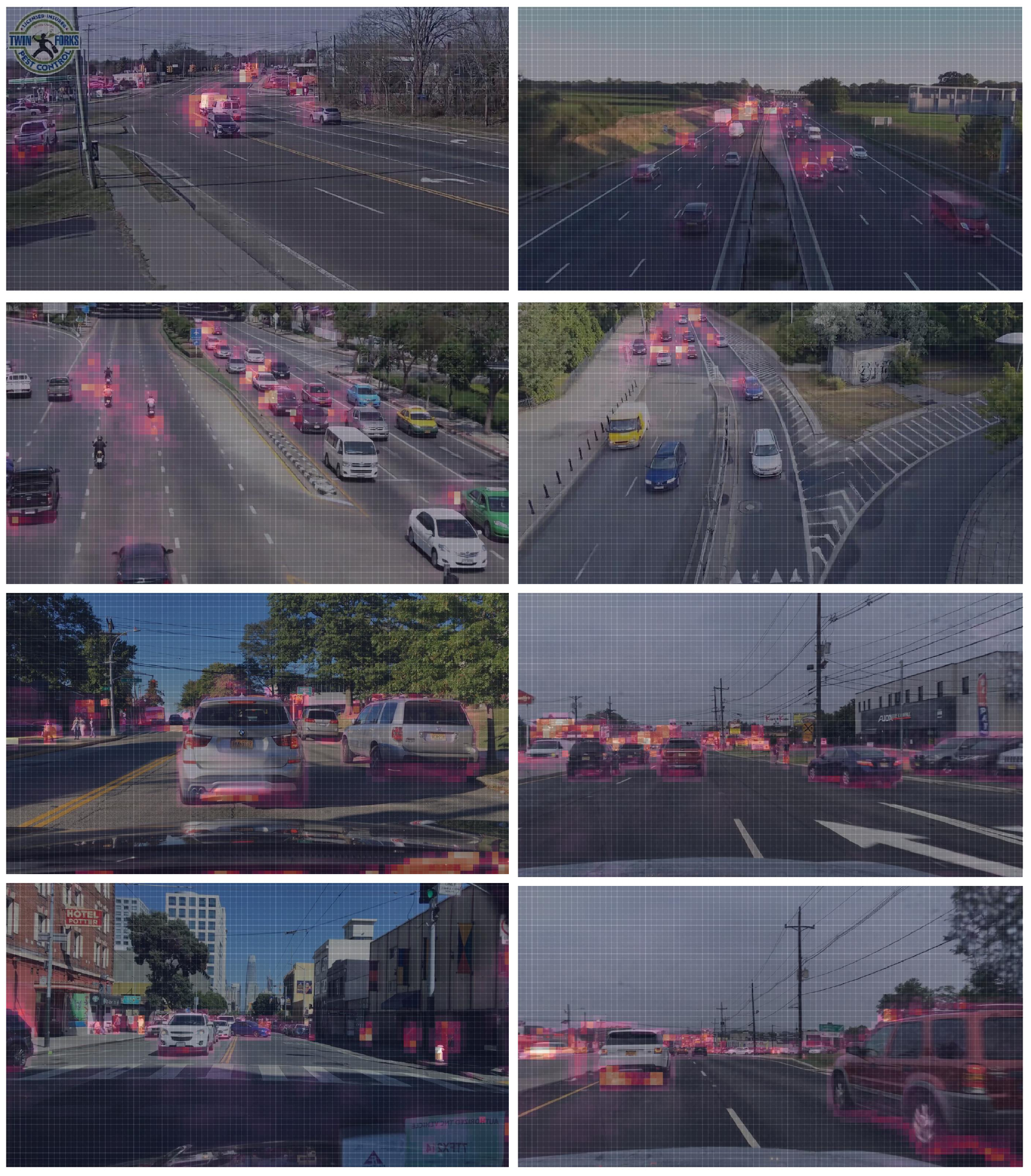}
    \caption{Representative mask*.}
    \label{fig:mask star}
\end{figure*}


\end{document}